\documentclass{elsart}
\usepackage{graphicx}
\usepackage{bm}
\usepackage{amsmath}

\begin{document}
\begin{frontmatter}

\title{Conductance distribution in quasi-one-dimensional disordered
quantum wires}

\author{K. A. Muttalib},
\address{Physics Department, University of
Florida, P.O. Box 118440, Gainesville,
FL 32611-8440, USA.}
\author{P. W\"olfle} and \author{V. A. Gopar}
\address{Institut f\"ur Theorie
der Kondensierten Materie, Universit\"at Karlsruhe,
D-76128 Karlsruhe, Germany.}

\begin{abstract}

We develop a simple systematic method, valid for all strengths of
disorder, to obtain analytically the full distribution of
conductances $P(g)$ for a quasi one dimensional wire within the
model of non-interacting fermions. The method has been used in
\cite{mu-wo,gmw,mwgg} to predict sharp features in $P(g)$ near
$g=1$ and the existence of non-analyticity in the conductance
distribution in the insulating and crossover regimes, as well as
to show how $P(g)$ changes from Gaussian to log-normal behavior
as the disorder strength is increased. Here we provide many
details of the
method, including intermediate results that offer much insight
into the nature of the solutions. In addition, we show within the
same framework that while for metals $P(g)$ is a Gaussian around
$\langle g \rangle \gg 1$, there exists a log-normal tail for $g
\ll 1$, consistent with earlier field theory calculations. We also
obtain several other results that compare very well with available
exact results in the metallic and insulating regimes.
\end{abstract}

\end{frontmatter}

\section{Introduction}
The absence of self-averaging in mesoscopic disordered conductors
leads to important fluctuation effects in the electronic transport
properties \cite{altshuler}. In the metallic regime, this gives
rise to the universal conductance fluctuations \cite{altshuler1}.
In this regime, for a given strength of macroscopic disorder, the
probability distribution $P(g)$ of the dimensionless conductance
$g$ (in units of $e^2/h$), for all possible microscopic
distributions of randomness, is a Gaussian with a universal
variance {\it i.e.} its value does not depend on microscopic
details of the sample but depends on the symmetry of the system
only. With sufficiently large disorder however, the
moments of the conductance fluctuations can become of the same
order of magnitude as the average conductance, and therefore the
average value becomes insufficient in describing the statistical
properties of the conductor. In the extreme case of the deeply
insulating regime, $P(g)$ becomes log-normal, which means that it
has a very long tail so that its mean value is very different from
the most probable value.  Thus for all strengths of disorder
beyond the metallic regime, one must consider the full
distribution of conductances. In particular, a very broad
distribution may change the qualitative nature of the Anderson
transition, the metal to insulator transition at zero temperature
even in the absence of electron-electron interactions
\cite{mohanty}. The numerically obtained $P(g)$ at critical
disorder for the three dimensional (3D) Anderson transition is
indeed very broad and highly asymmetric \cite{markos1}. The
conductance at the integer quantum Hall transition is also
expected to have a very broad, almost flat, distribution
\cite{wang,plerou}.

There is no theoretical method currently available to obtain
directly the full distribution of conductances for all strengths
of disorder in arbitrary dimensions \cite{note}. Moments of the
distribution have been calculated in dimension $D=2+\epsilon$,
$\epsilon \ll 1$ \cite{akl} within the supersymmetric non-linear
sigma model framework \cite{efetov} but the critical distribution
obtained from the moments \cite{shapiro} in $D=2+\epsilon$ differ
qualitatively from the numerically obtained $P(g)$ in 3D. In the
present work we will consider the distribution of conductances for
the simpler case of a quasi one dimensional (quasi 1D) wire, for
which the width $W$ is  smaller than the elastic mean free path
$l$, and much smaller than its length $L$. Although such a system
has no phase transition, it has  well defined metallic and
insulating regimes, and a smooth crossover between them. The
distribution in the crossover regime, where the localization
length is of the order of the system size, should give a
reasonably good qualitative picture of the distribution near the
critical regime in higher dimensions, where the localization
length diverges and becomes of the order of the system size. Even
for this simple quasi 1D geometry, only the first two moments
of $P(g)$ have
been obtained for all strengths of disorder \cite{mirlin}, using
the non-linear sigma model. Numerical \cite{{plerou},{martin}} as well
as experimental \cite{poirier} results show
a broad asymmetric distribution qualitatively similar to 3D.

We will use a scattering approach in which the conductance is
given in terms of the transmission eigenvalues $T_i$ of a finite
conductor sandwiched between two ideal leads \cite{landauer}. The
finite width of the lead quantizes the perpendicular momenta into
$N$ values, providing $N$ channels of scattering and therefore $N$
number of transmission levels. The dimensionless conductance $g$
is simply given by the total transmission probability
\begin{equation}
g=\sum_{i=1}^N T_i.
\end{equation}
The statistics of these levels as a function of the length $L \gg
l$ of the conductor is described by the
Dorokhov-Mello-Pereyra-Kumar (DMPK) equation \cite{dmpk}.
This equation has been shown \cite{frahm} to be equivalent to the
non-linear sigma model \cite{efetov} obtained from the microscopic
tight binding Anderson model, where electrons hop to neighboring
sites that have random energies chosen from a  uniform
distribution. An advantage of using the DMPK equation is that its solution
{\it i.e.} the joint probability
distribution (jpd) of all the transmission eigenvalues is known
\cite{beenakker} and since the conductance is simply related to these
eigenvalues, it is possible to write down
the full distribution of conductances. The major problem
is that the jpd involves $N$ constrained integrals for the $N$
eigenvalues, and that we would be interested in the case of a large
number of
transmission eigenvalues.

We have developed a simple systematic method, valid for {\it all
strengths of disorder}, to evaluate analytically the full
distribution of conductance for a quasi 1d wire, starting from the
solution of the DMPK equation. For simplicity, we restrict our
calculation mostly to the unitary symmetry case where time
reversal symmetry is broken, {\it e.g.} by the application of a magnetic
field. The method is valid for other symmetry classes also, and we
will discuss extension of the calculation to these cases as well.
As a check of the essential framework, we will show analytically
that the method reproduces the known exact results for the  mean,
variance as well as the distribution of the conductances in both
the metallic and insulating limits. We will also compute the
leading correction to the mean conductance in the metallic regime,
which agrees very well with the exact value obtained in
\cite{mirlin}. Since the leading order correction to the variance
vanishes for the unitary symmetry class, we will evaluate it for
the other two symmetry classes; again the results agree very well
with the exact results. Also, it has already been shown in
\cite{gmw} that the mean and the variance as a function of
disorder obtained within the method agrees qualitatively well with
exact results for all strengths of disorder including the
crossover regime.

Some of the novel features of the distribution obtained within our
method have been reported earlier
\cite{mu-wo,gmw,mwgg}. Here we discuss details which
provide insights about the rare but allowed configurations of the
transmission eigenvalues obtained from the method and allow us to
study these features. In particular, we provide details that led
to the prediction of a `one-sided' log-normal distribution near
the crossover regime in \cite{mu-wo}, the changes in the shape of
the distribution across the crossover regime obtained in
\cite{gmw} as well as the non-analyticity in the distribution near
$g=1$ predicted in \cite{mwgg}. We also show for the first time
within DMPK that while for metals the distribution is a Gaussian
around $\langle g \rangle \gg 1$, there exists a long non-Gaussian
tail for $g \ll 1$, consistent with earlier field theoretic
calculations \cite{akl}.

The paper is organized as follows. In section \ref{DMPK} we review
the DMPK equation and its general solution. In section
\ref{saddlepoint} we introduce the basic strategy of our new
approach which is used to evaluate $P(g)$ from the jpd of
the $T_i$ for all strengths of disorder in terms of certain saddle
point free energies. Our method is based on separating out an
appropriate number of discrete eigenvalues and treating the rest
as a continuum. For simplicity, we start with separating out only
one eigenvalue in section \ref{saddlepoint}. In section
\ref{free-energy} we obtain explicit forms for the density of the
continuum and various free energy terms. In this section we also
discuss some of the special features in the density and their
implications for the distribution. In section \ref{metallic} we
focus on the metallic regime and obtain several analytical results
and compare them with known exact results. We also obtain the
non-Gaussian tail of the distribution. In section 
\ref{insulating} we obtain approximate analytic expressions for
the distribution in the insulating and crossover regimes based on
the separation of one eigenvalue. In section \ref{2eigen} we
extend our formulation by separating out two eigenvalues, which
have been used to obtain the non-analyticity in $P(g)$ near $g=1$
and also to evaluate the distribution across the crossover regime.
In section \ref{betadif2} we  briefly discuss the case of
other symmetries, and obtain corrections to the mean and variance
of $g$ and compare them with known exact results. We provide
a summary and discuss conclusions in section \ref{summary}.

\section{\label{DMPK}Brief review of DMPK equation}

For a quasi 1D wire with $N$ channels, as shown in Fig. 1, the
probability distribution $p(\lambda)$ of the
variables $\lambda_i=(1-T_i)/T_i$, where $T_i$ are the
transmission eigenvalues, satisfies the well known DMPK equation
\cite{dmpk}
\begin{equation}
l\frac{\partial p}{\partial L}=\frac{2}{N+1}
\frac{1}{J(\lambda)}\sum_a\frac{\partial}
{\partial \lambda_a}\left[\lambda_a(1+\lambda_a)
J(\lambda)\frac{\partial p(\lambda)}
{\partial \lambda_a}\right],
\end{equation}
where $l$ is the mean free path,
$J(\lambda)=\prod_{i<j}^{N}|\lambda_i -\lambda_j|^\beta$ and the
parameter $\beta$ depends on the symmetry
of the ensemble \cite{dyson}: $\beta=1$ for orthogonal symmetry
with time reversal invariance, $\beta=2$ for unitary symmetry
where time reversal symmetry is broken by e.g. the application of
a magnetic field, and $\beta=4$ for symplectic symmetry where e.g.
spin-orbit scattering is present.
\begin{figure}
\begin{center}
\includegraphics[width=0.5\textheight]{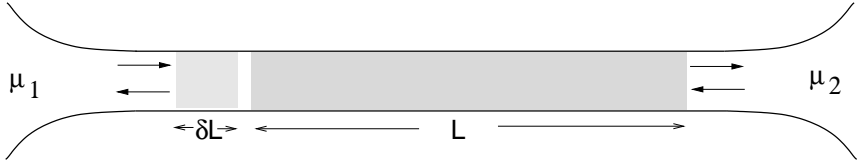}
\end{center}
\caption{Quasi-one dimensional disordered wire attached to two
electron reservoirs $\mu_1, \mu_2$. The DMPK equation describe the
evolution of
the joint distribution of the transmission eigenvalues $T_i$ when system size
L is increased {\it e.g.} by $\delta$L}
\end{figure}
The DMPK equation has been solved for all three
values of $\beta$ \cite{beenakker,caselle}. The general solution is
complicated, but simplify in the metallic and insulating regimes.
In these two regimes it can be written in the general form
\begin{equation}
\label{poflambda}
p(\lambda)=\frac{1}{Z}\exp[-\beta H(\lambda)],
\end{equation}
where $Z=\int\prod_i d\lambda_i \exp[-\beta H]$ is a
normalizing factor independent of $\lambda_i$, and
$H(\lambda)$ may be interpreted as the Hamiltonian
function of $N$ classical charges at positions $\lambda_i$,
interacting with a two-body repulsive interaction but
confined by a  one-body confinement potential, given as
\begin{equation}
H=\sum_{i<j}^{N}u(\lambda_i,\lambda_j)
+\sum_{i}^{N}V(\lambda_i).
\end{equation}

We will  use the variables $\lambda_i=\sinh^2 x_i $, in terms of
which the solution in the metallic regime $1 \ll L/l \ll N$  is
given by
\begin{eqnarray}
&&u(x_i,x_j)=-\frac{1}{2}\ln \vert \sinh^2 x_i-\sinh^2 x_j\vert
-\frac{1}{2}\ln \vert x^2_i - x^2_j\vert,  \\
&&V(x_i)=\frac{\gamma l}{2L\beta}x^2_i-\frac{1}{2\beta} \ln \vert
x_i \sinh 2x_i\vert. \nonumber.
\end{eqnarray}
Here $\gamma=\beta N + 2 - \beta$. In the insulating regime the
solution is given by
\begin{eqnarray}
&&u_{ins}(x_i,x_j)=-\frac{1}{2}\ln \vert \sinh^2 x_i-\sinh^2
x_j\vert
-\frac{1}{\beta}\ln \vert x^2_i - x^2_j\vert,  \\
&&V(x_i)=\frac{\gamma l}{2L\beta}x^2_i-\frac{1}{2\beta} \ln \vert
x_i \sinh 2x_i\vert-\frac{1}{2\beta}\ln x_i \nonumber.
\end{eqnarray}
The parameters $N$, $L$ and $l$ appear in the above expressions in
the combinations $L/l$ and $\Gamma=Nl/L$. We will consider the
limit where both $N$ and $L$ approach infinity, keeping $\Gamma$
fixed.  The limit $\Gamma \gg 1$ will correspond to the metallic
regime while $\Gamma \ll 1$ will correspond to the localized
regime. Note that for $\beta=2$, the solution in the insulating
regime differs from the solution in the metallic regime only by a
logarithmic term in the one-body potential.

In terms of the variables $\lambda_i$ or $x_i$, the dimensionless
conductance is given by
\begin{equation}
g=\sum_{i}^{N}\frac{1}{1+\lambda_i} =\sum_{i}^{N}\frac{1}{\cosh^2
x_i}.
\end{equation}
In the insulating regime, $\langle g \rangle \ll 1$ and $x_i \gg
1$. In this limit both $\ln x_i$ and $\ln|x^2_i-x^2_j|$ terms in
Eq. (6) are negligible compared to the other terms. Since these are
the only terms that differ from Eq. (5), we can assume that the
metallic expression (5) is valid for all regimes.

In the metallic regime $\langle g \rangle \gg 1$, the $\lambda_i$
are very close to each other so that a continuum description can
be used with a density of $\lambda$ points finite between zero and
an upper cutoff given by the normalization condition. Strictly
speaking, Eq. (\ref{poflambda}) is valid only for $x^2 >
2/\beta\Gamma$ \cite{caselle}, but this approximation works well
in the metallic  regime. Within  a global maximum entropy ansatz,
such a continuum approximation was used \cite{muttalib} to show
the relevance of a Wigner-Dyson like random matrix ensemble
\cite{mehta} for the transmission levels. The density of levels
obtained from the DMPK solution within such a continuum
approximation yields the correct value of the mean and the
variance in the metallic regime \cite{beenakker}, known from
diagrammatic perturbation theory \cite{altshuler1}. In the deeply
insulating regime on the other hand, all $\lambda_i$ are
exponentially large and separated exponentially from each other,
and conductance is dominated by the smallest eigenvalue. This
approximation yields the log-normal distribution of the
conductance in the deeply insulating regime
\cite{pichard,beenakker}. It is clear however that none of the
above descriptions can be used in the crossover regime, where the
smallest eigenvalue is neither zero, nor exponentially large. Our
approach combines the essential features of the two descriptions
and provides a simple and systematic procedure to study the
conductance distribution $P(g)$ at all strengths of disorder.


\section{\label{saddlepoint}Generalized saddle point approximation for
$P(g)$}

Given the joint probability distribution $p(\lambda)$ and the
definition of $g$ Eq. (7), the distribution of conductance can be
written as
\begin{equation}
P(g)=\frac{1}{Z}\int_{-\infty}^{\infty}
\frac{d\tau}{2\pi}\int_{0}^{\infty}\prod_{i=1}^{N}
d\lambda_i  \exp\left[i\tau \left(g-\sum_{i}^{N}\frac{1}
{1+\lambda_i}\right)\right]p(\lambda).
\end{equation}
For later use, we define a ``Free energy''
\begin{equation}
F(\lambda; \tau)=\sum_{i}^{N}\frac{i\tau}{1+\lambda_i}+\beta H
\end{equation}
such that the distribution of $g$ can be written as
\begin{equation}
P(g)=\frac{1}{Z}\int_{-\infty}^{\infty} \frac{d\tau
}{2\pi}e^{i\tau g}\int_{0}^{\infty}\prod_{i=1}^{N} d\lambda_i
\exp\left[-F(\lambda; \tau)\right].
\end{equation}
This defines a statistical mechanical problem of $N$ classical
charges in one dimension, at an inverse temperature $\beta$ and
described by the free energy $F$, which includes a source term
proportional to $\tau$. The basic idea of our method is to first
separate out an appropriate number of the lowest eigenvalues, and
treat the rest as a continuum by introducing a density for those
eigenvalues. In \cite{mu-wo} a single eigenvalue
was separated out, which was sufficient to obtain the qualitative
nature of $P(g)$ in the metallic regime, in the crossover region
on the insulating side as well as deep in the insulating regime.
In \cite{gmw,mwgg}, the importance of separating at least two
eigenvalues in order to study the crossover region quantitatively
was emphasized. However here we will consider in detail the case
of separating out one eigenvalue in the metallic regime since it
allows us to obtain many of our results analytically, which
provides much insight into the nature of the solutions, and two
eigenvalues in the crossover and insulating regimes where it
becomes essential. The `density' of the continuum, including the
source term, is obtained following Dyson \cite{dyson} by
minimizing the free energy functional, which then yields a saddle
point free energy. The qualitative nature of the density gives
important information about the problem and will
not only provide guidelines for appropriate approximations, but
will also suggest novel possibilities in the distribution.
We will see that at least for $\beta=2$, the $\tau$
dependence of $F(\lambda;\tau)$ is only quadratic, so the integral
over $\tau$ in (10) can be done exactly. The evaluation of $P(g)$
is then reduced to two integrals in the metallic regime and three
integrals in the crossover and insulating regimes, over all
possible positions of the separated discrete levels, as well as
over the beginning of the continuum. These integrals can then be
done approximately analytically in some limiting cases to obtain
qualitative results, or numerically to obtain quantitative results
for comparison with direct numerical simulations. While we will
mainly restrict our considerations to the case $\beta=2$ only, we
would point out the results for other values of $\beta$ whenever
they can be obtained without too much difficulty.

In the following, we will discuss the steps in more detail.

\subsection{Separation of the lowest eigenvalue}

We start by separating out only the lowest eigenvalue from the
rest. This  will be sufficient to study the metallic regime. In
Sec. 7, we will show how to add a second discrete level, and write
down the rules to extend the one-discrete level formulae to the
two-discrete level case. This will keep the discussion of the
basic framework simpler.

In the variable $x_i$, we get
\begin{equation}
\label{Hsum}
H=H_1+\sum_{1<i<j} u(x_i,x_j)+\sum_{i>1} V(x_i),
\end{equation}
where
\begin{equation}
H_1=\sum_{i>1} u(x_i,x_1)+V(x_1).
\end{equation}
The idea is to treat the rest as a continuum, beginning at
$x_2 > x_1$.

\subsection{Continuum approximation for the remaining eigenvalues}

Before going to a continuum description, we note that $u(x_i,x_j)$
can be written as
\begin{equation}
u(x_i,x_j)= u^{\prime}(x_i-x_j)+u^{\prime}(x_i+x_j) ,
\end{equation}
\begin{displaymath}
u^{\prime}(x)=-\frac{1}{2}\ln (2x \sinh x) . \nonumber
\end{displaymath}
The continuum description will then include the $x_i=x_j$
terms in $u^{\prime}(x_i+x_j)$ which are not included in the
discrete part. In order to get rid of these terms we
rewrite the sum Eq. (\ref{Hsum}) as
\begin{eqnarray}
H&=&H_{1} \nonumber \\
&+&\sum_{i>1} V(x_i) +\frac{1}{2}\left[ \sum_{1< i\ne j}
u^{\prime}(x_i-x_j) +\sum_{i,j>1} u^{\prime}(x_i+x_j) -\sum_{i>1}
u^{\prime}(2x_i)\right].
\end{eqnarray}
The $x_i=x_j$ terms in $u^{\prime}(x_i-x_j)$ are infinite, and
will not make a difference in the continuum limit. Using in
addition the expressions for $V(x_i)$ and $u^{\prime}_2(2x_i)$,
we obtain
\begin{equation}
H=H_{1}+\frac{1}{2}\sum_{i,j>1} u(x_i,x_j)
+\sum_{i>1}\left[V_{\Gamma}(x_i)+2\alpha \ln (x_i \sinh 2x_i)
\right],
\end{equation}
where
\begin{equation}
V_{\Gamma}(x_i)=\frac{\Gamma}{2} x^2_i \hspace{0.7cm}
\alpha=\frac{1}{8} \left( 1- \frac{2}{\beta} \right)
\end{equation}
Thus for $\beta=2$, the term proportional to $\alpha$ drops
out.

The continuum approximation can now be made, giving for $\beta=2:$
\begin{eqnarray}
\label{Hcont}
H(x_1,x_2; \sigma(x))&=&V(x_1)+\int_{x_2}^b dx \sigma(x)u(x,x_1)
+\int_{x_2}^b dx \sigma(x)V_{\Gamma}(x) \nonumber \\
&+&\frac{1}{2}\int_{x_2}^b dx \int_{x_2}^b dx^{\prime}
\sigma(x)u(x,x^{\prime})\sigma(x^{\prime})
\end{eqnarray}
where
\begin{equation}
V(x_1)=\frac{\Gamma}{2}x^2_1
-\frac{1}{2\beta}\ln (x_1 \sinh 2x_1)
\end{equation}
and the `density' $\sigma(x)$ has to be obtained in a consistent
way, subject to the condition that it is positive everywhere. The
upper limit $b$ in principle should take care of the normalization condition
\begin{equation}
\int_{x_2}^b \sigma (x) dx = N-1.
\end{equation}
However, we will be interested in the $N\rightarrow\infty$ limit, and since
the contributions to the conductance from large eigenvalues are
negligible, we will not worry about enforcing the normalization condition and
replace the upper limit by $\infty$.
The Free energy can now be written in the continuum
approximation as
\begin{equation}
F(x_1,x_2;\sigma(x))=\beta H(x_1,x_2; \sigma(x))
+\frac{i\tau}{\cosh^2 x_1}
+i\tau\int_{x_2}^{\infty}\frac{\sigma(x)}{\cosh^2 x} dx
\end{equation}
and $P(g)$ is given by a functional integral on
the generalized density $\sigma(x)$:
\begin{equation}
\label{pofg}
P(g)=\frac{1}{Z}\int_{-\infty}^{\infty}
\frac{d\tau}{2\pi}e^{i\tau g}\int_{0}^{\infty}dx_1
\int_{x_1}^{\infty}dx_2 \int D[\sigma(x)]
\exp[-F(x_1,x_2;\sigma(x))].
\end{equation}

\subsection{Integral equation for the saddle point density}

We now minimize the free energy functional by taking a
functional derivative with respect to $\sigma(x)$ keeping
$x_1$ and $x_2$ fixed:
\begin{equation}
\frac{\delta F(x_1,x_2;\sigma(x))}{\delta \sigma(x)}=0.
\end{equation}
This gives an integral equation for the saddle point density.
We will see later
that because the saddle point density goes to zero close
to the origin, we will need to consider fluctuation
corrections to the saddle point function.

Following Dyson (generalized for the non-logarithmic
interaction case in \cite{beenakker}), we have the saddle point equation
for $\sigma(x)$:
\begin{equation}
\label{spe}
-\int_{x_2}^{\infty} dy \sigma (y)u(x,y) +4
\alpha\left[\ln \sigma (x)+\delta u(x,y)+\frac{l}{L}x^2\right] =V_{tot}(x)+
{\rm const,}
\end{equation}
where $\delta u(x,y)=u(x,y)+\ln\vert x-y\vert$. Here $V_{tot}$ now
includes
\begin{equation}
V_{tot}(x)=V_{\Gamma}(x)+u(x,x_1)+\frac{i\tau}{\cosh^2 x}.
\end{equation}
Again the term proportional to $\alpha$ drops out for
$\beta=2$.

\subsection{The shift approximation}
We define
\begin{equation}
u_1=\ln \vert \sinh^2 x-\sinh^2 y \vert; \;\;\;
u_2=\ln \vert x^2-y^2 \vert,
\end{equation}
so that $u(x,y)=-\frac{1}{2}(u_1+u_2)$.
Then Eq. (\ref{spe}) for $\beta=2$ becomes
\begin{equation}
\int_{x_2}^{\infty} dy \sigma (y)[u_1(x,y)+u_2(x,y)]
=2V_{tot}(x)+ {\rm const};  \;\;\;\; x > x_2 .
\end{equation}
In the limit $x_2\rightarrow 0$, the density can be extended
symmetrically to negative values which makes it possible to
invert the kernel and obtain $\sigma(x)$. Here the finite
lower limit poses difficulty. A shift in the
variables $x$ and $y$ to make the lower limit zero does not help,
because the kernels are not translationally invariant. However in the
variable $\lambda$, the  kernel
$u_1(\lambda-\lambda^{\prime})=\ln \vert \lambda-\lambda^{\prime} \vert$
is translationally invariant while
$u_2(\lambda,\lambda^{\prime})
=\ln \vert {\rm arsinh} \sqrt{\lambda}
- {\rm arsinh} \sqrt{\lambda^{\prime}} \vert$
remains invariant for a metal ($\lambda \ll 1$),
and is negligible in the insulating regime ($\lambda \gg 1$)
compared to $u_1$. This suggests the following shift
approximation to leading order, to which systematic
corrections can be made:

(i) Write the integral equation in variable $\lambda$
\begin{equation}
\label{inteq}
\int_{\lambda_2}^{\infty}d\lambda^{\prime}
[u_1(\lambda-\lambda^{\prime})+u_2(\lambda,\lambda^{\prime})]
\rho(\lambda^{\prime})
=2V_{tot}(\lambda).
\end{equation}
where $\rho(\lambda^{\prime})d\lambda^{\prime}=\sigma(x)dx$.
Now shift to $\eta=\lambda-\lambda_2,$ and
$\eta^{\prime}=\lambda^{\prime}-\lambda_2$ to make the lower
limit zero:
\begin{equation}
\int_{0}^{\infty}d\eta^{\prime}
[u_1(\eta-\eta^{\prime})+u_2(\eta+\lambda_2,
\eta^{\prime}+\lambda_2)]
\rho(\eta^{\prime}+\lambda_2)
=2V_{tot}(\eta+\lambda_2).
\end{equation}

(ii) Write $u_2(\eta+\lambda_2,\eta^{\prime}+\lambda_2)
=u_2(\eta, \eta^{\prime})+\Delta u_2$ where
\begin{equation}
\Delta u_2(\eta, \eta^{\prime};\lambda_2)
=\ln \frac{{\rm arsinh}^2 \sqrt{\eta+\lambda_2}
-{\rm arsinh}^2 \sqrt{\eta^{\prime}+\lambda_2}}
{{\rm arsinh}^2 \sqrt{\eta}
-{\rm arsinh}^2 \sqrt{\eta^{\prime}}}.
\end{equation}
For both $\eta \gg \lambda_2$ and $\eta \ll \lambda_2$, $\Delta
u_2$ is negligible. In the metallic
regime, the leading term of $\Delta u_2$ is a small correction
linear in $\lambda_2$. So, for $\lambda_2 \ll 1$ it is given by 
\begin{equation}
\Delta u_2(\eta,\eta^{\prime};\lambda_2)=\frac{\lambda_2}{{\rm arsinh}^2
\sqrt{\eta}
-{\rm arsinh}^2 \sqrt{\eta^{\prime}}}\left[
\frac{{\rm arsinh} \sqrt{\eta}}{\sqrt{\eta(1+\eta)}}
-\frac{{\rm arsinh} \sqrt{\eta^{\prime}}}
{\sqrt{\eta^{\prime}(1+\eta^{\prime})}}
\right] .
\end{equation}

In the opposite limit {\it i.e.} in the
insulating regime ($\lambda_2 \gg 1$), the entire term $u_2$ is negligible
compared to $u_1$. Therefore we will keep the leading term of
$\Delta u_2$ as a small correction
to the free energy in the metallic regime only. Its
contribution to the saddle
point density itself would be negligible. In this spirit,
we make the approximation:
\begin{equation}
\int_0^{\infty}d\eta^{\prime}
\Delta u_2(\eta, \eta^{\prime};\lambda_2)
\rho(\eta^{\prime}+\lambda_2)
\approx\int_0^{\infty}d\eta^{\prime}
\Delta u_2(\eta, \eta^{\prime};\lambda_2)
\rho_{sp}(\eta^{\prime}+\lambda_2),
\end{equation}
where the saddle point density $\rho_{sp}$ is the density obtained
from the saddle point integral equation in the absence of the
$\Delta u_2$ term. Then the integral equation (\ref{inteq}) can be
rewritten with the $\Delta u_2$ term taken to the right hand side
to add to the $V_{tot}$, giving
\begin{equation}
\int_{0}^{\infty}d\eta^{\prime}
[u_1(\eta-\eta^{\prime})+u_2(\eta,\eta^{\prime})]
\rho_{sp}(\eta^{\prime}+\lambda_2)
=2V_{eff}(\eta;\lambda_2),
\end{equation}
where the effective potential $V_{eff}$ is defined as
$V_{eff}(\eta;\lambda_2)=V_{tot}(\eta+\lambda_2)-V_2(\eta;\lambda_2)$,
with $V_{tot}(x)$ defined in (24) and
\begin{equation}
\label{V2}
V_2(\eta;\lambda_2)=\frac{1}{2}\int_0^{\infty}d\eta^{\prime}
\Delta u_2(\eta, \eta^{\prime};\lambda_2)
\rho_{sp}(\eta^{\prime}+\lambda_2).
\end{equation}
In the crossover and insulating regimes, we will take $V_2=0$,
since $\Delta u_2$ is negligible in these regimes.

(iii) Now change the variables in the saddle point equation (32):
\begin{equation}
\sinh^2 t=\eta=\lambda-\lambda_2,\;\;\;
\sinh^2 s=\eta^{\prime}=\lambda^{\prime}-\lambda_2
\end{equation}
\begin{equation}
\label{spd1}
\int_{0}^{\infty}ds
\left[\ln \vert \sinh^2 t-\sinh^2 s \vert+\ln \vert t^2-s^2 \vert
\right]\sigma(s)
=2V_{eff}(\eta(t)+\lambda_2).
\end{equation}
where we have defined
\begin{equation}
\sigma(s)=\rho_{sp}(\eta^{\prime}
+\lambda_2)\frac{d\eta^{\prime}}{ds} =2\sinh s\cosh s \;
\rho_{sp}(\eta^{\prime}+\lambda_2).
\end{equation}

Defining $\sigma(-s)=\sigma(s)$ we extend the density
symmetrically to negative $s$ and using the relation: $\sinh^2
t-\sinh^2 s=\sinh(t-s)\sinh(t+s)$, the integral (\ref{spd1}) can
also be extended to $-\infty$, giving
\begin{equation}
\int_{-\infty}^{\infty}ds
[\ln \vert \sinh(t-s) \vert+\ln \vert t-s \vert]\sigma(s)
=2V_{eff}(\eta(t)+\lambda_2).
\end{equation}
Since $\eta(-s)=\eta(s)$, this is true for all $s$.

(iv) Taking a derivative on both sides of the integral
equation, we get
\begin{equation}
\int_{-\infty}^{\infty}ds K(t-s)\sigma(s)
=2\frac{d}{dt}V_{eff}(\eta (t)+\lambda_2)
\end{equation}
where the kernel
\begin{equation}
K(t-s)=\coth (t-s)+\frac{1}{t-s}.
\end{equation}
This kernel $K(t)$ can be inverted to give the saddle point
density
\begin{equation}
\sigma(t)=2\int_{-\infty}^{\infty} ds K^{-1}(t-s)\frac{d}{ds}
V_{eff}(\eta(s)+\lambda_2).
\end{equation}
Using Fourier transform, we invert the kernels
$K_1(t)=1/t$ and  $K_2(t)=\coth t$ to obtain $K_1(q)=i\pi sign (q)$
and  $K_2(q)=i\pi\coth (\pi q/2)$.
Then using $\sigma (q)=[K_1(q)+K_2(q)]^{-1} \; 2V^{\prime}_{eff}(q)$
we write
\begin{equation}
K^{-1}(t)=\int_{-\infty}^{\infty}\frac{dq}{2\pi}
\frac{e^{-iqt}}{K_1(q)+K_2(q)}
=-\frac{1}{\pi^2}\int_0^{\infty}dq \sin (qt)\frac{1}{2}(1-e^{-\pi q}).
\end{equation}
Finally, introducing Eq. (41) into (40), the saddle point density is given
by
\begin{equation}
\sigma(t)=\frac{2}{\pi^2}\int_0^{\infty}
dq \cos (qt) (1-e^{-\pi q})
\int_0^{\infty}ds \sin (qs) \frac{d}{ds}
V_{eff}(\eta(s)+\lambda_2) ,
\end{equation}
using the fact that $V_{eff}$ is even in $s$.
Note that for $\beta \ne 2$, there would be additional terms
in $V_{eff}$ proportional to the parameter $\alpha$.

\subsection{ Contributions to the Free energy}

From the saddle point density, we obtain the saddle point Free
energy by using the right hand side of the integral equation
(\ref{spe}) to eliminate one of the two integrals of the last term
in Eq. (\ref{Hcont}). Substituting this result into Eq. (20) and
changing to the $\lambda$ variable (since shift has to be made in
the $\lambda$ variable), we write the Free energy as
\begin{equation}
F_{sp}(\lambda_1,\lambda_2)= \frac{\beta}{2}
\int_{\lambda_2}^{\infty}d \lambda
V_{tot}(\lambda)\rho_{sp}(\lambda)+\beta V(\lambda_1).
\end{equation}
Shifting by $\lambda_2$ to make the lower limit zero, and
then going to the variable $t,s$ and using the expression for
the density, we get
\begin{equation}
F_{sp}(\lambda_1,\lambda_2)=\frac{\beta}{2}\int_0^{\infty}dt
V_{tot}(\eta(t)+\lambda_2)\sigma(t)+\beta V(\lambda_1).
\end{equation}
We will write
\begin{equation}
V_{tot}(\eta (t)+\lambda_2)=V_{\Gamma}+V_{u_1}+V_{u_2}+V_z
\end{equation}
with corresponding densities $ \sigma(t)
=\sigma_{\Gamma}+\sigma_{u_1}+\sigma_{u_2}+\sigma_z $ (Note that
for $\beta \ne 2$, $V_{tot}$ has to be modified, Sec. 8), where
\begin{eqnarray}
V_{\Gamma}&=&\frac{\Gamma}{2} \left(\sinh^{-1} \sqrt{\sinh^2 t+
\lambda_2}\right)^2 , \nonumber \\
V_{u_1}&=&-\frac{1}{2}\ln \left(\sinh^2 t+\lambda_2-\lambda_1\right) ,
\nonumber \\
V_{u_2}&=&-\frac{1}{2}\ln \left[\left(\sinh^{-1}\sqrt{\sinh^2 t
+\lambda_2}\right)^2 -\left(\sinh^{-1}\sqrt{\lambda_1}\right)^2\right] ,
\nonumber \\
V_z&=&\frac{z/\beta}{\cosh^2 t+\lambda_2}; \;\;\;\;z=i\tau.
\end{eqnarray}
In addition, we define a density component $\sigma_2$,
generated by $V_2$ as defined in (\ref{V2}). So the total free
energy would be a sum of all combinations of partial free energies
\begin{equation}
F=\sum_{a,b}F_{ab}+F_1
\end{equation}
where $a,b$ runs through the four indices $\Gamma,u_1,u_2,z$ as
defined above and $b$ runs through an additional index $2$, and
\begin{equation}
F_{ab}=\frac{\beta}{2}\int_0^{\infty}dt
V_a(\eta(t)+\lambda_2)\sigma_b(t),
\end{equation}
with the partial density
\begin{equation}
\sigma_b(t)=\frac{2}{\pi^2}\int_0^{\infty}
dq \cos (qt) (1-e^{-\pi q})
\int_0^{\infty}ds \sin (qs) \frac{d}{ds}
V_b(\eta(s)+\lambda_2).
\end{equation}
Here $F_1=\beta V(\lambda_1)$. By doing a partial integration for
both the $t,s$ integrals, we see that the boundary terms are
independent of $\lambda$, giving $F_{ab}=F_{ba}$. Also, we
note that we need to obtain the partial densities in order to calculate
$V_2$.

\subsection{The action for $P(g)$}

Since $V_{eff}$ and therefore $\sigma$ are
both linear in $\tau$ for $\beta =2$, the free energy is quadratic in
$\tau$ and can be written in the form
\begin{equation}
F_{sp}=F^0+(i\tau)F^{\prime}+\frac{(i\tau)^2}{2} F^{\prime\prime}
\end{equation}
and the integral over $\tau$ in (\ref{pofg}) can  be done
exactly. The result is
\begin{equation}
P(g)=\frac{1}{Z}\int_{0}^{\infty}d\lambda_1
\int_{\lambda_1}^{\infty}d\lambda_2 e^{-S},
\end{equation}
where the saddle point action $S$ is given by
\begin{equation}
S=-\frac{1}{2F^{\prime\prime}} (g-F^{\prime})^2+F^0.
\end{equation}

Note that the quadratic form in (50) is no longer true for 
$\beta \ne 2$, because of  
the term $\ln(\sigma(x))$ in (\ref{spe}) which contains $\tau$
inside the logarithm. However, we will see later that the leading
terms in the free energy are still quadratic in $\tau$, and will
allow us to calculate  the leading corrections to $\langle g \rangle$ and
$var(g)$.

\subsection{Mean and variance of $g$ as function of $\Gamma$}
We can obtain the moments of $P(g)$ directly from the free energy
instead of obtaining the full $P(g)$ first. The nth moment of
conductance is given by
\begin{equation}
\langle g^n\rangle = \int_0^{\infty} dg\;
g^n P(g)/\int_0^{\infty} dg P(g),
\end{equation}
defining the integrals $I_k$ as
\begin{equation}
I_k(x_1,x_2)=\int_0^{\infty}dg g^k e^{-\frac{(g-F^{\prime})^2}
{2\vert F^{\prime\prime}\vert}}
\end{equation}
which can be done analytically:  
$I_0=\sqrt{2\vert F^{\prime\prime}\vert}f_0(Q) $;
$I_1=\vert F^{\prime\prime}\vert e^{-Q^2}
+F^{\prime}I_0$;
$I_2=-F^{\prime 2}I_0+2F^{\prime}I_1+
(2\vert F^{\prime\prime}\vert)^{3/2}f_2(Q)$,
where
$f_0(Q)=(\sqrt{\pi}/2)[1+{\rm Erf(Q)}]$ with {\rm Erf(Q)} the Error function, 
$f_2(Q)=(f_0(Q)-Qe^{-Q^2})/2$, and
$Q=F^{\prime}/\sqrt{2\vert F^{\prime\prime}\vert}$ . Defining
\begin{equation}
J_k=\int_0^{\infty}dx_1\int_{x_2}^{\infty}dx_2 e^{-F^0}I_k(x_1,x_2),
\end{equation}
we have that the average and variance of $g$ can be expressed as
\begin{equation}
\langle g \rangle =J_1/J_0; \;\;\; var (g)=J_2/J_0-(J_1/J_0)^2.
\end{equation}


\section{\label{free-energy}Density and Free energy}

In order to evaluate the free energy terms explicitly, we first
need the five components of the total density, given by Eq. (49).
While $\sigma_z$ and $\sigma_{u_1}$ can be evaluated exactly
analytically, the others require approximations. In particular
$\sigma_2$ can be evaluated only after the potential $V_2$ has
been obtained from the other partial densities, and $V_2$ itself
involves  approximations. In principle, one could evaluate all the
free energy terms directly numerically without any further
approximations. This will involve at most triple integrals, Eqs.
(48) and (49). More importantly, we will see that the qualitative
nature of the various partial densities provide much insight into
the nature of the problem and not only provides guidelines for
appropriate approximations in various regimes, but also suggests
novel possibilities in the distribution. We therefore will
evaluate the density integrals analytically, albeit approximately
when necessary. Here we list the results for the different
$\sigma$'s; details of the calculations are given
in Appendix A.

For $\sigma_z(t)$ and $\sigma_{u_1}(t)$ the results are exact while
for
$\sigma_{u_2}(t)$ and $\sigma_{\Gamma}(t)$ good approximations can
be obtained analytically by modelling the derivatives of the
potentials $V_\Gamma$ and $V_{u_2}$ as
\begin{equation}
V'_{\Gamma}(s)\approx\Gamma s\left[1-\frac{a_4}{\cosh 2s+
\cosh \delta_1}\right],
\end{equation}
where $a_4$ and $\delta_1$ are chosen to match the
exact form at $s\rightarrow 0$ and at $s=x_2$, and
\begin{equation}
V'_{u_2}(s)\approx-s\frac{s^2+s^2_m}{(s^2+s^2_m)^2+4c^2s^2}
\left[1-\frac{a s^2_m}{s^2+s^2_m}\right],
\end{equation}
with $a$ and $c$ obtained by matching $V'_{u_2}(s)$ at $s=0$ and
$s=s_m$, $s_m$ being chosen to follow the maxima which occurs at
$s=x_d$ for $x_d\rightarrow 0$ and at $s\approx x_2$ for $x^2_d
\gg 1$, where $x^2_d=x_2^2-x_1^2$. Using these approximations we
list the partial densities:
\begin{subequations}
\begin{equation}
\sigma_z(t)=-\frac{z/\beta}{\sqrt{\lambda_2(1+\lambda_2)}}
\left[\frac{x_2+t}{((x_2+t)^2+(\pi/2)^2)^2}+
\frac{x_2-t}{((x_2-t)^2+(\pi/2)^2)^2}\right], 
\end{equation}
\begin{equation}
\sigma_{u_1}(t)= -\frac{1}{\pi}\left[
\frac{(\pi-a_0)/2}{(\pi-a_0)^2/4+t^2}
+\frac{(\pi+a_0)/2}{(\pi+a_0)^2/4+t^2}\right],
\end{equation}
\begin{eqnarray}
\sigma_{u_2}(t)&\approx&-\frac{A_+}{2\pi\beta_1}
\left(\frac{\beta_1+c}{t^2+(\beta_1+c)^2}-
\frac{\pi+\beta_1+c}{t^2+(\pi+\beta_1+c)^2}\right) \nonumber \\
       &&+ \frac{A_-}{2\pi\beta_1}\left(\frac{\beta_1-c}{t^2+(\beta_1-c)^2}-
\frac{\pi+\beta_1-c}{t^2+(\pi+\beta_1-c)^2}\right) ,
\end{eqnarray}
\begin{equation}
\sigma_{\Gamma}(t)\approx\Gamma\left(
1+\frac{a_4\delta_1}{4\sinh\delta_1} \frac{\pi^2 +
\delta^2_1}{(t^2+(\pi^2-\delta_1)/4)^2+(\pi\delta_1/2)^2}-
\frac{a_4}{\cosh 2t + \cosh \delta_1}   \right)
\end{equation}
\end{subequations}
with the following parameters:
\begin{eqnarray}
a_0=\cos^{-1}[2(\lambda_2-\lambda_1)-1]; \;\;\;\;
A_{\pm}=\beta_1\pm c
+a s_m^2/2c; \;\;\;\; a=1-\frac{s_m^2}{x_2 p_2};  \nonumber \\
\beta_1^2=c^2+s_m^2; \;\;\;\;s_m=x_2\sinh^{-1}p_2; \;\;\;\;\;
p_2=\frac{\sinh2x_2}{2x_2^2}(x_2^2-x_1^2).   \nonumber
\end{eqnarray}

\subsection{Positivity of the density}

Both $\sigma_{u_1}$ and $\sigma_{u_2}$ diverge with negative
coefficients in the limit $t=0$ (i.e. $\lambda=\lambda_2$) and
$\lambda_2-\lambda_1\rightarrow 0$. This means that the total density
(without the
source term $\sigma_z$) given by
\begin{equation}
\sigma_{tot}(t)=\sigma_{\Gamma}+\sigma_{u_1}+\sigma_{u_2}
\end{equation}
will always become negative near $t=0$ in this limit. Negative
total density for any $t$ is clearly an unacceptable solution. To
determine what to do with these negative density solutions, we
need to understand why our saddle point equations produce such
solutions. Both $\sigma_{u_1}$ and $\sigma_{u_2}$ are
contributions from the repulsive interaction terms. The total
density becomes negative when  $\lambda_2-\lambda_1$ becomes less
than certain value depending on $\Gamma$. Thus the negative total
density corresponds to the unacceptable configurations of
$\lambda_2$ too close to $\lambda_1$, which are very costly in
energy due to the mutual repulsion of the eigenvalues. It is
therefore clear that such solutions should be discarded. In fact,
positivity of the total density can be guaranteed simply by
imposing a minimum on the allowed values of $\lambda_2-\lambda_1$,
such that $\sigma_{tot} (t=0) \ge 0$. In the metallic limit
$\lambda_2 \ll 1$, and the condition of positivity becomes
\begin{equation}
x^2_2-x^2_1 \ge \left( \frac{2}{\pi \Gamma} \right)^2,
\end{equation}
while in the insulating limit $\lambda_2 \gg \lambda_1 \gg 1$ we
have
\begin{equation}
x_2 - x_1 \ge 1/2\Gamma.
\end{equation}
Figure 2a shows the total density $\sigma_{tot} (t)$ for $x_1=0.1$
and three different values of $x_2$ in the metallic regime; it
shows how the negative part of the density for small $t$ gradually
disappears with increasing $x_2$, such that the density can always
be made positive for all $t$ by choosing the appropriate lower cut
off $x_r$ for $x_2$.
\begin{figure}
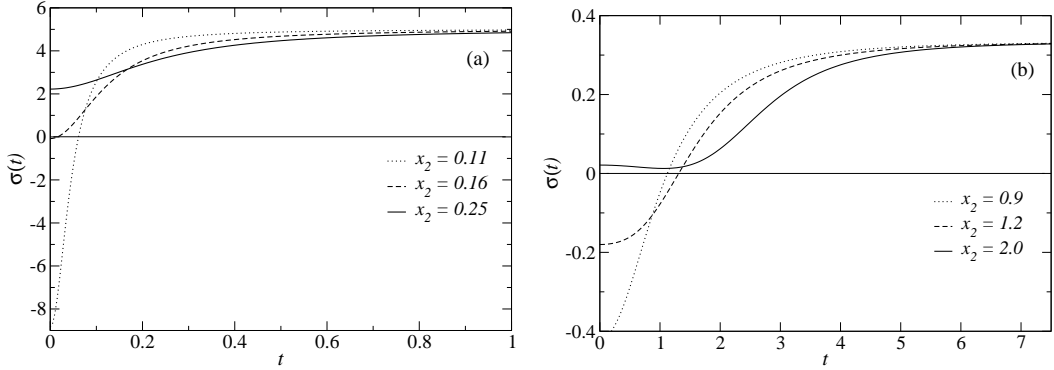

\vspace*{.25in}
\begin{center}
\includegraphics[width=0.298\textheight]{mu-wo-go-fig2a.eps}
\hspace{0.1cm}
\includegraphics[width=0.298\textheight]{mu-wo-go-fig2b.eps}
\end{center}
\caption{Total density $\sigma_{tot}(t)$ as a function of $t$ (a)
in the metallic regime $\Gamma=5$ with fixed $x_1=0.1$ for three
different values of $x_2$ and (b) in the insulating regime
$\Gamma=1/3$ with fixed $x_1=0.5$, for three different values of
$x_2$, showing that the negative part of the density gradually
disappears with increasing $x_2$. Note that in the metallic
regime, $\sigma (t)$ saturates at $\Gamma$ for large $t$.}
\end{figure}
In all our analytical evaluations on the metallic side we will use
the lower cut off (61). Note that since $x_2-x_1 \ll 1$ in the
metallic limit, the restriction (61) also imposes a condition
$\Gamma \gg 2/\pi$ for which a positive density metallic solution
can be obtained. This gives a crude estimate that the metallic
region ends and the crossover region begins at $\Gamma \approx
1/2$. Similarly, Fig. 2b shows $\sigma_{tot} (t)$ for $x_1=0.5$
and three different values of $x_2$ in the insulating regime.
\begin{figure}
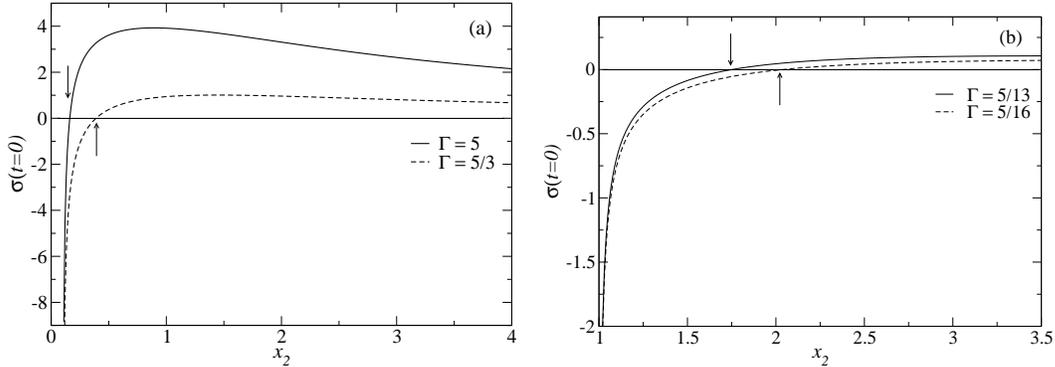

\vspace*{.25in}
\begin{center}
\includegraphics[width=0.298\textheight]{mu-wo-go-fig3a.eps}
\hspace{0.1cm}
\includegraphics[width=0.298\textheight]{mu-wo-go-fig3b.eps}
\end{center}
\caption{$\sigma_{tot}(t=0)$ as a function of $x_2$ for two
different values of $\Gamma$: (a) in the metallic regime, for a
given fixed $x_1=0.1$ and (b) in the insulating regime, for a
given fixed $x_1=1.0$,  and two different values of $\Gamma$,
showing how the cutoff for the positivity of density depends on
disorder.}
\end{figure}
In figure 3a, $\sigma (t=0)$ is plotted as a function
of $x_2$ for two
different values of $\Gamma$ in the metallic regime for a fixed
$x_1=0.1$, while Fig. 3b shows the same in the insulating regime for
a fixed $x_1=1.0$. From this figure 3b we can see how the increase
in disorder (decrease
in $\Gamma$) increases the cutoff for $x_2$ that ensures the
positivity of the density. Fig. 4 shows several numerically
evaluated roots $x_2=x_r$ of $\sigma(t=0)$. For comparison with Eq.
(61), we plot the scaled quantities $(\Gamma x_{r})^2$ as a
function of $(\Gamma x_1)^2$ which, according to Eq. (61), should
be a straight line in the metallic regime. Figure 4 shows that the
metallic limit agrees very well with Eq. (61), and that the same
equation provides a good estimate and an upper bound in the
crossover and insulating regimes as well. Nevertheless, in our
numerical evaluations, for each $x_1$, we will find the value of
$x_r$ numerically by evaluating the total density at $t=0$.
Because of the cutoff in $x_2$ at $x_r$, $P(g)$ will be given by
\begin{equation}
P(g)=\int_{0}^{\infty}dx_1
\int_{x_r}^{\infty}dx_2 e^{-S(x_1,x_2)}.
\end{equation}
Note that since we used the variable $x$ instead of $\lambda$, the
action $S(x_1,x_2)$ will contain appropriate Jacobian terms.
\begin{figure}
\vspace*{.25in}
\begin{center}
\includegraphics[width=0.35\textheight]{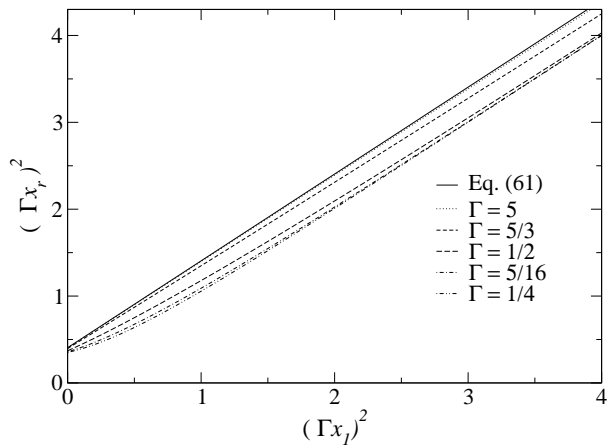}
\end{center}
\caption{Values of $x_2=x_r$ for which $\sigma_{tot}(t=0)$ scaled
by $\Gamma$ plotted as a function of $(\Gamma x_1)^2$ for several
$\Gamma$ in the metallic, insulating and crossover regimes. Eq. 
(61) fits the metallic regime well, and is a good estimate and an
upper bound for the insulating and crossover regions.}
\end{figure}

The inequalities (61, 62) show that the separation between $x_1$
and $x_2$ increases with decreasing $\Gamma$. Since $\Gamma \gg 1$
describes metals and $\Gamma \ll 1$ describes insulators, this
gives rise to the well-known qualitative  picture that all
eigenvalues in metals are close to the origin while they are all
large and well-separated in the insulating regime. In addition,
Eqs. (61) and (62) suggest two more qualitatively distinct regions
involving rare but allowed configurations:

(i) $\Gamma \gg 1$, $x_1 \gg 1$, $x_2 \gg 1$, but $x_2\approx x_1$. We
will show in this paper  that even for metals, this gives rise to the
existence of long tails in $P(g)$
in the region $g \ll 1$ (Sec. 5).

(ii) $\Gamma \ll 1$, $x_1 \ll 1$, $x_2 \gg 1$. In this insulating
region a Gaussian cut off in the log-normal conductance
distribution for $g>1$ was found in \cite{mu-wo} (Sec. 
6).

Thus the qualitative features of the density provide us with much insight, not
only about various approximations and their limitations, but also suggesting
rare but allowed configurations giving rise to novel  possibilities in the
distribution.

\subsection{Fluctuation correction to the functional integral}

As we have seen, the generalized density $\sigma(t)$ goes to zero
at $t=0$. This means that the saddle point evaluation of the
functional integration over the density requires corrections. The
fluctuation correction will involve the integral
\begin{equation}
I_\sigma=\int D[\sigma]\exp\left[-\frac{1}{2}\int_0^{\infty}dt
\int_0^{\infty}ds \delta\sigma(t)u(t,s)
\delta\sigma(s)\right]
\end{equation}
where $u(s,t)$ includes the two body terms, and
$\delta\sigma=\sigma-\sigma_{sp}$. The result is in general of the
form $ I_{\sigma}\sim1/\sqrt{det\vert u(t,s) \vert}. $ This will
give rise to a contribution to the free energy of the form $
\delta F=+\frac{1}{2}\ln (det\vert u \vert). $ Although it should
be possible to evaluate the determinant numerically, we can
already argue that the correction is usually negligible (see
Appendix C). The fluctuation correction to the free energy is
relevant only in the extreme metallic regime and it is given by
\begin{equation}
\label{deltaF}
\delta F\approx\frac{1}{2}\ln \lambda_d.
\end{equation}


\subsection{The partial free energies}

The partial free energies related with $\sigma_z$ and
$\sigma_{u1}$ given by Eq. (48) are calculated exactly for any
regime. In particular, $F_{zz}$ is relevant for the
calculation of the exact variance in the metallic regime as well
as to see the qualitative change in the solutions for large and
small $x_2$:
\begin{equation}
F_{zz}=-\frac{z^2}{\beta}\frac{1}{\sinh^2 (2x_2)}
\left[\frac{1}{3}-\frac{1}{4x^2_2}+\frac{1}{\sinh^2 2x_2}\right]
\;\;\;.
\end{equation}
There are also two direct terms
\begin{equation}
F^0_1=\frac{\beta}{2}\Gamma x^2_1-\frac{1}{2}\ln (x_1\sinh
(2x_1)); \;\;\; F^z_1=\frac{z}{\cosh^2 x_1},
\end{equation}
and the contribution from the fluctuation correction to the
functional integral $\delta F$ (\ref{deltaF}) valid for $x_2 \ll
1$. Other free energy terms can be calculated within the same
approximations as used for the densities (see Appendix B for
details). In particular, the free energies can be obtained
analytically in the metallic and insulating regimes, keeping the
leading orders in $x_2$. On the other hand, we will see later that
an interesting crossover regime on the insulating side will be
determined by the case where $x_2 \gg 1$, but  $x_1$ can be
arbitrary. Free energy expressions in this regime can also be
obtained approximately analytically. Note also that these
expressions will be valid for $\beta \ne 2$ as well, for which
there will be additional $V_j$ terms and therefore additional free
energy terms. For quantitative results on $P(g)$, we will
integrate numerically Eq. (48) to obtain the free energy.

In the final saddle point action, we have written the total free
energy in the form given in (50). Remembering a factor of two for
symmetrical terms $F_{ab}=F_{ba}$ for $b\ne a$, we then have
\begin{eqnarray}
F^0&=&F_{\Gamma\Gamma}+F_{u_1 u_1}+F_{u_2 u_2}+F^0_1+\delta F
+2[F_{u_1 \Gamma}+F_{\Gamma u_2}+F^{\Gamma}_{2\Gamma}+F_{u_1 u_2}] , \\
F^{\prime}&=&\frac{1}{z}F^z_1+\frac{2}{z}[F_{\Gamma z}+F_{z u_1}+F_{z u_2}
+F_{z 2}+F^z_{2 \Gamma}] , \\
F^{\prime\prime}&=&\frac{2}{z^2}F_{zz}.
\end{eqnarray}
Given the free energies, the action $S$ determines the full distribution
$P(g)$. We will discuss several quantitative results
in the metallic and deeply insulating regimes as well as qualitative results
in the insulating side of the crossover regime analytically. The
distribution in the crossover regime has to be obtained numerically. 

\section{\label{metallic}The metallic regime}

The metallic regime is characterized by $x_2 \ll 1$ and the Free
energy terms in this regime can be obtained analytically by
expanding the potentials (46). Details of the calculations are
given In Appendix B.

\subsection{Free energy in the metallic limit:}

Using $x_1 < x_2 \ll 1$ and therefore
$\sqrt{\lambda_d}\approx \sqrt{x^2_2-x^2_1}$, the total free
energy in the metallic limit is given by
\begin{eqnarray}
F^0 &\approx& a_1\Gamma^2 x^2_2 +a_2 \ln (x^2_2-x^2_1)
-a_3 \Gamma \sqrt{x^2_2-x^2_1}-\ln x_1 ,  \\
F^{\prime}&\approx& \Gamma -b_1 \Gamma x^2_2 +b_2\sqrt{x^2_2-x^2_1}; \;\;\;\;
F^{\prime\prime}\approx-\frac{1}{15},
\end{eqnarray}
where $a_1=3\pi^2/8$, $a_2=3/2$, $a_3=2\pi$, $b_1=2/3+2\left(
8-2\zeta(2)-3\zeta(3)\right)/\pi^2$ and $b_2=32/\pi^3$, where
$\zeta(n)$ is the Riemann zeta function.

\subsection{Saddle point analysis in the deep metallic limit}

Since $x_2 \ll 1$, and $\Gamma \gg 1$, we will keep only the
dominant first term in $F'$ for the purpose of obtaining a saddle
point solution (we will see that $x_2$ is $O(1/\Gamma)$, so that
the neglected terms are smaller by a factor $1/\Gamma^2$ compared
to the leading term ). In terms of the scaled variables
$\mu \equiv \Gamma \sqrt{x^2_2-x^2_1}$ and $\nu\equiv  \Gamma x_1$, the free
energy in this limit is
\begin{equation}
F^0 \approx a_1(\mu^2+\nu^2) +2 a_2 \ln \mu
-a_3 \mu-\ln \nu -\ln \mu + \frac{1}{2} \ln (\mu^2+\nu^2) ,
\end{equation}
\begin{equation}
F^{\prime}\approx \Gamma ; \;\;\;\;
F^{\prime\prime}=-\frac{1}{15} .
\end{equation}
The last two terms in $F^0$ comes from the Jacobian of
transformation
$dx_2=\mu d\mu/(\Gamma\sqrt{\mu^2+\nu^2})$.
Using (52), the solutions to the saddle point equations
$\partial S/\partial \mu=\partial S/\partial \nu=0$
are given by
\begin{equation}
\frac{1}{\nu^2}=\frac{\mu a_3-2a_2+1}{\mu^2}; \;\;\;\;
\frac{1}{\mu^2+\nu^2}=-2a_1+\frac{1}{\nu^2} .
\end{equation}
Eliminating $\nu$ we obtain a cubic equation for $\mu$, whose real
solution turns out to be less than the cutoff due to the
positivity of the density. This means that the metallic solution
is dominated by the boundary value
$\mu=\Gamma\sqrt{x^2_2-x^2_1}=2/\pi$. The corresponding
$\nu=\mu/\sqrt{2}$. Thus the saddle point analysis gives both
$x_1$ and $x_2$ to be of order $1/\Gamma$.

\subsection{Correction to $\langle g \rangle $ in the
metallic limit}

The nth moment of $P(g)$ can be obtained from Eqs.
(54) and (55). The average conductance in the metallic limit is
$\Gamma$ with leading correction $\sim 1/\Gamma$, for $\beta=2$, with
a coefficient obtained exactly in
\cite{mirlin}. In order to compare with \cite{mirlin}, we write
$\langle g \rangle$ as
\begin{equation}
\langle g\rangle =\frac{ \int_{0}^{\infty}dx_1
\int_{x_r}^{\infty}dx_2
\int_0^{\infty} dg [(g-F^{\prime})+F^{\prime}]e^{-S}}
{\int_{0}^{\infty}dx_1
\int_{x_r}^{\infty}dx_2
\int_0^{\infty} dg e^{-S}}.
\end{equation}
The first integral over $g$ in the numerator can be done simply,
and we write
\begin{equation}
\int_0^{\infty} dg [(g-F^{\prime})+F^{\prime}]e^{-S}=
\vert F^{\prime\prime}\vert
e^{-\frac{F^{\prime 2}}{2\vert F^{\prime\prime}\vert }-F^0}
+F^{\prime}\int_0^{\infty} dg e^{-\frac{(g-F^{\prime})^2}
{2\vert F^{\prime\prime}\vert }-F^0}.
\end{equation}
In the metallic regime, $F^{\prime}\sim \Gamma \gg 1$, and $\vert
F^{\prime\prime}\vert=1/15 \ll 1$; the first term in (77) is exponentially
small and we neglect it. The numerator and the denominator in (76) then
have the same integral over $g$. For $F^{\prime}\sim \Gamma \gg
1$, this integral  depends on $\vert F^{\prime\prime}\vert$ only. Since
$\vert F^{\prime\prime}\vert$ is a constant, it
cancels out from the ratio and we have
\begin{equation}
\langle g\rangle \approx \frac{ \int_{0}^{\infty}dx_1
\int_{x_r}^{\infty}dx_2 F^{\prime}e^{-F^0}}
{\int_{0}^{\infty}dx_1
\int_{x_r}^{\infty}dx_2 e^{-F^0}}.
\end{equation}
The leading term in $F^{\prime}$ is $\Gamma$, independent of
$x_1$, $x_2$, so the leading term for the average
conductance is simply $\Gamma$. Scaling $\bar{x}_i=x_i\Gamma$,
we write
\begin{equation}
\langle g \rangle =\Gamma -\frac{\eta}{\Gamma}
\end{equation}
where
\begin{equation}
\eta\approx \frac{\int_{0}^{\infty}d\bar{x}_1
\int_{\bar{x}_r}^{\infty}d\bar{x}_2
\left[b_1 \bar{x}^{2}_2-b_2\sqrt{\bar{x}^{2}_2-\bar{x}^{2}_1}\right]
e^{-\bar{F}^{0}}}
{\int_{0}^{\infty}d\bar{x}_1
\int_{\bar{x}_r}^{\infty}d\bar{x}_2 e^{-\bar{F}^{0}}},
\end{equation}
\begin{equation}
\bar{F}^{0}=a_1\bar{x}^{2}_2
+a_2\ln (\bar{x}^{2}_2-\bar{x}^{2}_1)
-a_3\sqrt{\bar{x}^{2}_2-\bar{x}^{2}_1}-\ln \bar{x}_1,
\end{equation}
and $\bar{x}_r=\sqrt{\bar{x}^{2}_1+(2/\pi)^2}$. Note that
while the free energy expressions are valid only for $x_i \ll 1$,
the scaled variables $\bar{x}_i$ can be very large. However, the
results become insensitive to the upper limit beyond 4. Setting
the upper limit at 4 and doing the integral numerically we get
\begin{equation}
\eta=.027
\end{equation}
compared to the exact result $\eta=1/45\approx.022$. This shows
that our results are good in the metallic limit up to order
$1/\Gamma$. This is consistent with the fact that $x_i \sim
1/\Gamma,  (i=1,2)$, and that we kept only terms up to order $x_i$
and $\Gamma x^2_i$ in Eqs. (71, 72).

\subsection{Variance in the metallic limit}

Using $g^2=(g-F^{\prime})^2+2gF^{\prime}-F^{\prime 2}$, and the
above approximation that $F^{\prime 2}/2|F^{\prime\prime}| \gg 1$
with $F^{\prime\prime}$ constant, we can write the second moment,
following the same procedure as above, as $\langle g^2
\rangle=\langle |F^{\prime\prime}|+F^{\prime 2}\rangle$. Then the
variance is simply
\begin{equation}
var (g)=\langle g^2 \rangle-(\langle g \rangle)^2=|F^{\prime\prime}|
=\frac{1}{15},
\end{equation}
which is the correct variance in the metallic limit. We do not
calculate the correction to the variance for $\beta=2$, because the
leading correction
is of order $1/\Gamma^2$, and our approximate free energy expressions
are not good up to this order in the metallic regime. However, the
leading correction for $\beta \ne 2$ is of order $1/\Gamma$, and we will
see later that our approximations yield values very close to  the
exact result.

More generally, we can evaluate the integrals $J_k$ defined in
(55) numerically and obtain the average and variance of $g$ as a
function of the disorder parameter $\Gamma$. Results of these
calculations have been reported in \cite{gmw}. They are in very
good agreement with the exact results in the metallic region. 

\subsection{Conductance distribution in the metallic limit}

The saddle point analysis of the metallic regime gives both $x_1$
and $x_2$ of order $1/\Gamma$, which justifies keeping only the
leading term in $F'$ in Eq. (72), namely $F'=\Gamma$, deep in the
metallic regime. Then the distribution $P(g)$ becomes trivial,
because the $g$-dependent part can be taken out of the integrals
over $x_1$ and $x_2$ and the integrals only give the normalization
factor. We obtain the known Gaussian distribution:
\begin{equation}
P(g)\propto e^{-\frac{15}{2}(g-\Gamma)^2},
\end{equation}
with mean $\langle g \rangle =\Gamma$ and $var(g)=1/15$. Using
Eqs. (71, 72), $P(g)$ can be directly obtained by numerical
integration. As an example, Fig. 5  shows the
result of the numerical integration for $\Gamma=2$, compared with
Eq. (84) appropriately normalized.
\begin{figure}
\begin{center}
\includegraphics[width=0.4\textheight]{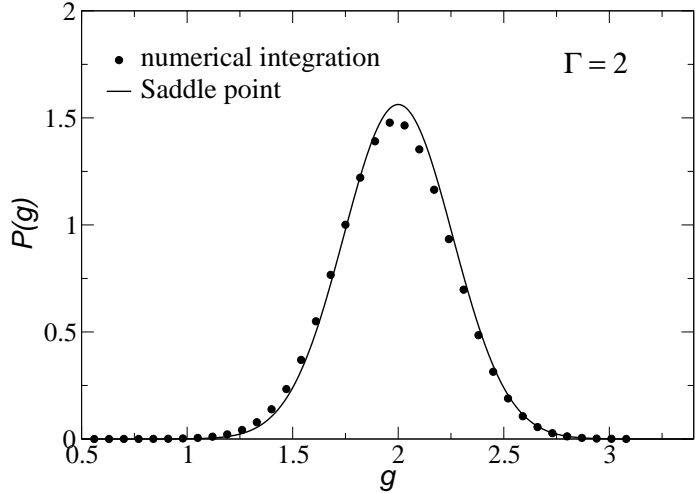}
\end{center}
\caption{Gaussian distribution of the conductance in the metallic
regime numerically obtained from Eqs. (71) and (72) for
$\Gamma=2$, compared with the saddle point result Eq. (84).}
\end{figure}
\subsection{$P(g)$ for $g\ll 1$ in the metallic limit}

An important question for the metallic distribution is whether
there are long (non-Gaussian) tails for $g\ll 1$. These will
signal rare events (localized states) dominating the tail of the
distribution, as suggested by field-theory models \cite{akl}. With
appropriate approximations of our free energy, we can analytically
consider this region qualitatively. We will be interested only in
$g \ll 1$, which can happen only if $x_1 \gg 1$ and $x_2 \gg 1$.
However, metal means large $\Gamma$, and the positivity of the
density requires that $x^2_2-x^2_1 > 4/\pi^2\Gamma^2$, Eq. (61).
This means that $x_2$ can be very close to $x_1$ (in contrast, for
the insulating case, $\Gamma \ll 1$, and $x_2 \gg x_1$).

The free energy in the limit $x_2 \gg 1$ can be written as
\begin{equation}
F^0\approx 2\Gamma^2 x^2_2 - 6\Gamma x_2
+\Gamma x^2_1-\frac{1}{2} \ln (x_1 \sinh (2x_1));
\end{equation}
\begin{equation}
F^{\prime}\approx \frac{1}{\cosh^2 x_1}+\frac{2\Gamma}{\sinh 2x_2};\;\;\;
F^{\prime\prime}\approx -\frac{4}{3}e^{-4x_2}.
\end{equation}
The saddle point solution is given by
\begin{equation}
\frac{\partial S}{\partial x_1}=\frac{2(g-F^{\prime })}
{2\vert F^{\prime\prime}\vert}
\left(-\frac{\partial F^{\prime}}{\partial x_1}\right)
+\frac{\partial F^0}{\partial x_1}=0.
\end{equation}
Since the quantity $1/\vert F^{\prime\prime}\vert$ is exponentially
large, the saddle point
solution is given simply by putting its coefficient equal to zero, i.e.
$g=F^{\prime}+O(e^{-4x_2})$.

For both $x_1 \gg 1$ and $x_2 \gg 1$ but $x_2\approx x_1$ and
$\Gamma \gg 1$, the second term in $F'$ can become the dominant
term (the first term dominates for $x_2 \gg x_1 \gg 1$ and $\Gamma
\ll 1$ which will describe the insulating region, and also for
$x_1 \ll 1$ and $\Gamma \approx 1$ which will describe the
crossover region). Then the saddle point solution for $x_2$ gives
$g=2\Gamma/\sinh 2x_2$. The leading order solution is $ x_2\approx
\frac{1}{2}\ln(4\Gamma /g), $ leading to
\begin{equation}
P(g)\propto e^{-2\langle g \rangle^2\left(\ln\frac{4\langle g \rangle}{g}
-\frac{3}{2\langle g \rangle}\right)^2};
\;\;\; g \ll 1,\;\;\; \langle g \rangle \gg 1,
\end{equation}
where we have used the fact that in the metallic limit $\langle g
\rangle =\Gamma$. The fluctuation correction does not add any
further $g$ dependence.
\begin{figure}
\vspace*{.25in}
\begin{center}
\includegraphics[width=0.35\textheight]{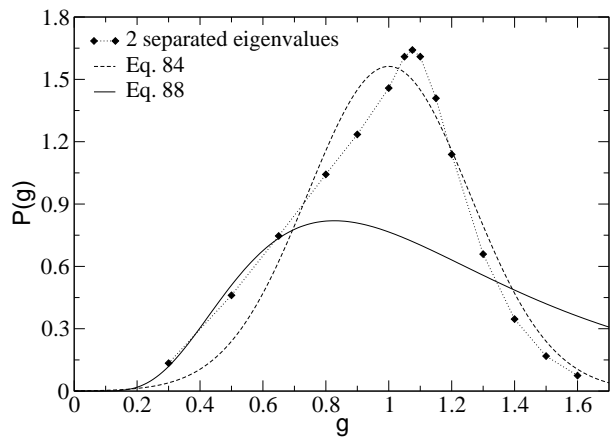}
\caption{$P(g)$ for $\Gamma=1.0$ obtained in [2] compared with
normalized (88) (solid line) and normalized (84) (dashed line).
The log-normal tail of the distribution for $g \ll 1$ is clearly
distinguishable from the Gaussian tail. Note that $\Gamma=1.0$ is
slightly beyond the metallic regime (a larger value would be
metallic) where (84) is not strictly valid. Nevertheless, it
clearly shows the Gaussian tail in the $g \gg 1$ limit as expected
for a metal but a log-normal tail in the $g \ll 1$ limit.}
\end{center}
\end{figure}
Figure 6 shows $P(g)$ given by  Eq. (88) compared with the
distribution of $g$ calculated from the two separated eigenvalue
case for $\Gamma =1.0$, as well as the Gaussian tail expected from
Eq. (84). Note that $\Gamma=1.0$ is slightly beyond the metallic
regime (a larger value of $\Gamma$ would be metallic).
Nevertheless it clearly shows the existence of a log-normal tail
in the $g \ll 1$ regime compared to a Gaussian tail in the $g \gg
1$ regime for the same value of $\Gamma$ close to the metallic
limit.


\section{\label{insulating}The insulating and crossover regimes within
the one separated eigenvalue approximation}

As emphasized in \cite{gmw,mwgg}, we need to isolate at least two
eigenvalues in order to study the insulating and the crossover
regimes. This we will do later. On the other hand, the
one-eigenvalue framework allows us to obtain many analytical
results, which helps us to better understand the qualitative
nature of the solutions. We will therefore expect our results in
this section to be valid only qualitatively. In particular, we
will point out why the one eigenvalue separation fails very close
to $g=1$ even qualitatively. In all other cases the numerical
evaluations with two separated eigenvalues improve the one
eigenvalue results quantitatively, without changing them
qualitatively.

\subsection{Free energy and saddle point solution in the insulating
limit}
In the limit $\Gamma \ll 1$ and $x_2 \gg x_1 \gg 1$, the free energy
simplifies to
\begin{equation}
F^0\approx 2\Gamma^2 x^2_2 - 6\Gamma x_2
+\Gamma x^2_1-x_1
\end{equation}
\begin{equation}
F^{\prime}\approx 4e^{-2x_1}; \;\;\;
F^{\prime\prime}\approx -\frac{4}{3}e^{-4x_2}.
\end{equation}
As before, since $x_2 \gg 1$, the saddle point solution is given by
putting $g=F^{\prime}+O(e^{-4x_2})=4e^{-2x_1}$, so that
$x_{1sp}=-\frac{1}{2}\ln u$, $u=g/4$.
The saddle point value of $x_2$ is obtained from
\begin{equation}
\frac{\partial S}{\partial x_2}=3e^{4x_2}(g-F^{\prime})^2
+\frac{\partial F^0}{\partial x_2} =\frac{\partial F^0}{\partial
x_2}+O(e^{-4x_2})=0.
\end{equation}
This gives
$x_{2sp}=3/2\Gamma$,
independent of $g$. As we will see (Eq. 95), since $x_{1sp}=-\ln u/2$,
$x_{1sp}\sim 1/2\Gamma$, so
$x_{2sp} \gg x_{1sp}$ for $\Gamma \ll 1$.

\subsection{Conductance distribution in the insulating limit}

Since $x_{2sp}$ is independent of $g$, it only contributes to the
normalization constant. The distribution of conductance in saddle
point approximation is determined by $x_1$ only,
\begin{equation}
P_{sp}(g)\propto e^{-(-x_1+\Gamma x^2_1)} \propto
e^{-\frac{\Gamma}{4}(\ln u+1/\Gamma)^2}.
\end{equation}
To this we need to add the fluctuation correction around the saddle
point $P(g) \propto e^{-S_{sp}}/\sqrt{\vert S^{\prime\prime}\vert}$,
where
\begin{equation}
S^{\prime\prime}=\frac{\partial^2 S}{\partial x^2_1}
=\frac{g-F^{\prime}}{\vert F^{\prime\prime}\vert}
\left(-\frac{\partial^2 F^{\prime}}{\partial x^2_1} \right)
+\frac{1}{\vert F^{\prime\prime}\vert}
\left(\frac{\partial F^{\prime}}{\partial x_1}\right)^2
+\frac{\partial^2 F^0}{\partial x^2_1}.
\end{equation}
The first term is zero at the saddle point. The third term is
negligible compared to the exponentially large second term, which gives
$
S^{\prime\prime}(x_{sp})\propto g^2.
$
The distribution of $\ln g$ can then be written as
\begin{equation}
P(\ln g)=P(g)\frac{dg}{d\ln g}=g P(g)
\propto e^{-\frac{\Gamma}{4}(\ln (g/4)+1/\Gamma)^2}.
\end{equation}
This is a log-normal distribution with mean and variance given by
\begin{equation}
\langle -\ln (g/4) \rangle =\frac{1}{\Gamma};\;\;\;
var(\ln (g/4))=\frac{2}{\Gamma},
\end{equation}
which agrees with known results.

\subsection{Mean and variance of $g$ as function of $\Gamma$}

In the insulating limit, the density is zero, and the calculation
of the mean and the variance becomes simple. In the crossover
regime, one can use the full free energy to numerically obtain the
mean and the variance using (54-56). These two moments of $P(g)$
have been reported earlier in \cite{gmw} and compared with exact
result. They fail in the crossover regime, but the two separated
eigenvalue approximation substantially improves the agreement with
exact results both in the insulating and in the crossover regimes.

\subsection{Distribution in the crossover regime}

To obtain quantitative results in the crossover regime, we will
need to obtain the free energy terms by numerically evaluating Eq.
(51). However, the regime $x_2 \gg 1$, $x_1 \ll 1$, corresponds to
$\Gamma$\raisebox{-0.6ex}{$\stackrel{<}{\sim}$}1 which
characterizes the crossover regime from the insulating side. For
this case, the free energy contributions are the same as in Eqs.
(89,90). This time, however, the fact that $x_2 \gg x_1$ makes the
first term in $F'$ dominate. The results have been discussed in
\cite{mu-wo}, where a `one-sided log-normal distribution' with
sharp features in $P(g)$ at $g=1$ was predicted. Here as an
example, we show in Fig. 7 the distribution $P(g)$ for
$\Gamma=5/12$ calculated by using the results in \cite{mu-wo}. On
the other hand, note that the region $\Gamma$
\raisebox{-0.6ex}{$\stackrel{>}{\sim}$}1, which corresponds to the
crossover regime from the metallic side, is characterized by $x_2
\sim 1$, $x_1 \ll 1$. Our approximate free energies are not valid
in this regime.

The saddle point solution of $x_1$ is again given by
$g=F^{\prime}$ and we have $\cosh x_1=\frac{1}{\sqrt{g}}$. Note
that since $\cosh x_1 \ge 1$, the boundary of the saddle point
solution is given by $g \le 1$. It is the existence of this
boundary that makes the current solution fail very close to $g=1$.
The two separated eigenvalue framework reveals the existence of a
non-analyticity in the distribution near this point \cite{mwgg}.
\begin{figure}
\begin{center}
\includegraphics[width=0.4\textheight]{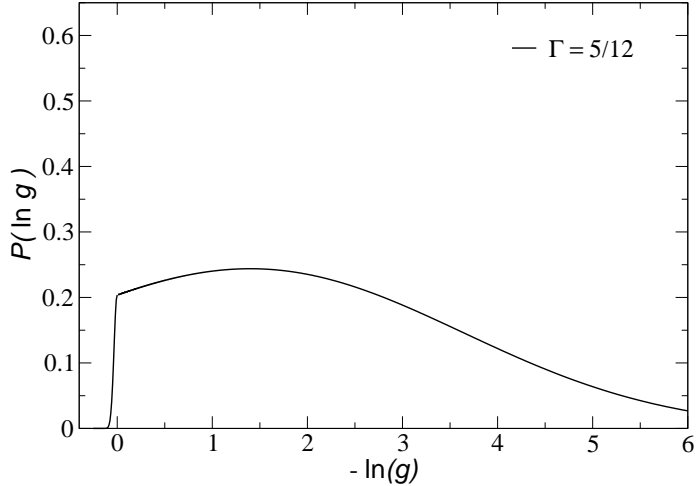}
\end{center}
\caption{The `one-sided log-normal distribution' of the conductance in the
crossover regime from the insulating side, where sharp features in
the distribution
were found around g=1 \cite{mu-wo}.}
\end{figure}

\section{\label{2eigen}Separating out two eigenvalues}

As mentioned earlier, the two separated eigenvalue case has been
treated in \cite{gmw,mwgg}. Here we give a summary for
completeness. Separating out 2 eigenvalues simply means that
equation (12) for $H_1$ is now replaced by
\begin{equation}
H_{1,2}=\sum_{i \ge 3} u(x_i,x_1)+\sum_{i \ge 3} u(x_i,x_2)+u(x_1,x_2)+
V(x_1)+V(x_2).
\end{equation}
The continuum approximation (17) now becomes
\begin{eqnarray}
&&H(x_1,x_2,x_3; \sigma(x))=V(x_1)+V(x_2)+u(x_1,x_2)+\int_{x_3}^b dx
\sigma(x)u(x,x_1) \nonumber \\ &&+\int_{x_3}^b dx \sigma(x)u(x,x_2)
+\frac{1}{2}\int_{x_3}^b dx \int_{x_3}^b dx^{\prime}
\sigma(x)u(x,x^{\prime})\sigma(x^{\prime})
+\int_{x_2}^b dx \sigma(x)V_{\Gamma}(x), \nonumber \\
\end{eqnarray}
while Eq. (20) for the free energy is now
\begin{equation}
F(x_1,x_2,x_3;\sigma(x))=\beta H(x_1,x_2,x_3; \sigma(x))
+\frac{i\tau}{\cosh^2 x_1}+\frac{i\tau}{\cosh^2 x_2}
+i\tau\int_{x_3}^{\infty}\frac{\sigma(x)}{\cosh^2 x},
\end{equation}
and the distribution $P(g)$ becomes
\begin{equation}
P(g)=\frac{1}{Z}\int_{-\infty}^{\infty}
\frac{d\tau}{2\pi}\e^{i\tau g}\int_{0}^{\infty}dx_1
\int_{x_1}^{\infty}dx_2 \int_{x_2}^{\infty}dx_3 \int D[\sigma(x)]
\e^{-F(x_1,x_2,x_3;\sigma(x))}.
\end{equation}
The shift parameter in (28) is now $\lambda_3$ and $V_{tot}$ Eq. (45)
has two additional terms
\begin{equation}
V_{tot}(\eta (t)+\lambda_3)=V_{\Gamma}+V_{u_{11}}+V_{u_{21}}+
+V_{u_{12}}+V_{u_{22}}+V_z
\end{equation}
with corresponding densities  $\sigma(t)
 =\sigma_{\Gamma}+\sigma_{u_{11}}+\sigma_{u_{21}}
+\sigma_{u_{12}}+\sigma_{u_{22}}+\sigma_z
$.
Here $V_{u_{11}}$ and $V_{u_{12}}$ are just the old $V_{u_1}$ and $V_{u_{2}}$
with
shift parameter $\lambda_2$ replaced by $\lambda_3$, (similarly for
$\sigma_{u_{11}}$ and
$\sigma_{u_{12}}$). The new terms
$V_{u_{21}}$ and $V_{u_{22}}$ can be written down following the
substitution
\begin{equation}
V_{u_{21}}=V_{u_1}(\lambda_2\rightarrow\lambda_3,
\lambda_1\rightarrow\lambda_2); \;\;\;
V_{u_{22}}=V_{u_2}(\lambda_2\rightarrow\lambda_3,
\lambda_1\rightarrow\lambda_2),
\end{equation}
and similarly for $\sigma_{u_{21}}$ and
$\sigma_{u_{22}}$.
In terms of the new saddle point densities and free energies, the
expression for $P(g) $ is similar to (51), with one extra
integral over $\lambda_3$:
\begin{equation}
P(g)=\frac{1}{Z}\int_{0}^{\infty}d\lambda_1
\int_{\lambda_1}^{\infty}d\lambda_2 \int_{\lambda_2}^{\infty}d\lambda_3
e^{-S}
\end{equation}
where the action $S$ has the same form as (52).
The mean and the variance can be obtained from the integrals $J_k$ of (56), now
given by
\begin{equation}
J_k=\int_0^{\infty}dx_1\int_{x_1}^{\infty}dx_2\int_{x_3}^{\infty}dx_2
e^{-F^0}I_k(x_1,x_2,x_3).
\end{equation}

The extra integral over $\lambda_3$ makes analytical calculations of $P(g)$
difficult. Even numerical evaluations  become involved. We will not only put
$V_2=0$ as discussed before, but also keep only the interaction between nearest
$\lambda$'s, neglecting in the potential terms the argument $\lambda_1$
compared to $\lambda_3$. Thus e.g. we set
\begin{equation}
V_{u_{11}}\approx -\frac{1}{2}\ln (\sinh^2 t+\lambda_3)
\end{equation}
Since $\lambda_3 \gg \lambda_1$ beyond the metallic regime, this is a valid
approximation in the regimes we are interested in.

The rest of the calculations are exactly as discussed for the one
eigenvalue case. For example, the density again becomes negative
for sufficiently small $x_3-x_2$. The positivity of the density is
guaranteed by numerically obtaining the minimum value of $x_3$ for
which the density is zero at its minimum.
\begin{figure}
\begin{center}
\includegraphics[width=0.4\textheight]{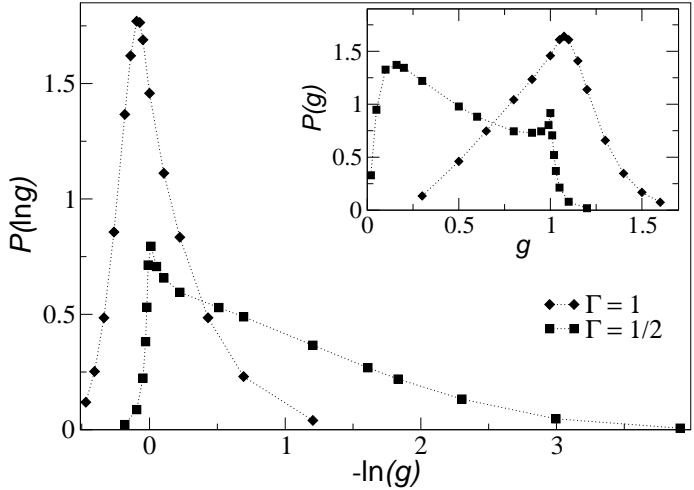}
\end{center}
\caption{Distribution of the conductance in the crossover regime.
A broad distribution is found for $\Gamma=1/2$ (squares);
eventually $P(\ln g)$ goes to the log-normal distribution in the
insulating regime. For $\Gamma =1$ (diamonds) $P(g)$ is closer to
the Gaussian distribution (inset, from \cite{gmw})  which is 
expected in the
metallic regime.}
\end{figure}
It is straightforward to repeat the evaluations for the mean and
the variance of $g$ as functions of $\Gamma$ within the two
separated eigenvalue framework in the crossover regime, and in the
insulating regime where the density is put equal to zero. These
results have been reported before \cite{gmw} and shows that they
are much better than the one eigenvalue case. In particular the
variance now follows the non trivial hump at $\Gamma=1/2$ in the
exact result \cite{mirlin}. Details of the results for $P(g)$ in
the crossover regime have also been reported in \cite{gmw}. They
compare well with numerical results available \cite{plerou},
including the direct Monte Carlo evaluation of Eq. (8)
\cite{spanishgroup}. As an example of the results obtained by
separating out two eigenvalues, we show in figure 8 the
distribution of the conductance as a function of $\ln g$ and $g$
(inset) in the crossover region. For $\Gamma=1/2$,
a broad distribution is observed, as reported in numerical
studies. On the other hand, $P(\ln g)$ goes, eventually in the
insulating regime, to a log-normal distribution (as the one shown
in Fig. 7). For $\Gamma=1$, which is closer to the metallic
regime, the profile of $P(g)$ is close to a Gaussian distribution,
as expected.

Note that in our derivation for the distribution in the crossover
region within the one separated eigenvalue framework, the saddle
point solution existed only on one side of $g=1$ while the other
side was dominated by the boundary value. Separating out two
eigenvalues allows us to treat the case near $g=1$ with more
accuracy. As discussed in \cite{mwgg}, we obtain a discontinuity
in $P'(g)=dP(g)/dg$ at $g=1+\alpha$ where
$\alpha=1/\cosh^2(3/2\Gamma)$ with $P'\sim e^{-2/\Gamma}$ for
$g\geq 1+\alpha$ and  with an almost flat distribution for
$g<1+\alpha$. In figure 9 we show $P(g)$ and its derivative
(inset) around $g=1$ for two different values of disorder, both in
the insulating regime, illustrating the discontinuity close to
$g=1$ reported in \cite{mwgg} and observed in numerical analysis
\cite{markos1}.
\begin{figure}
\begin{center}
\includegraphics[width=0.4\textheight]{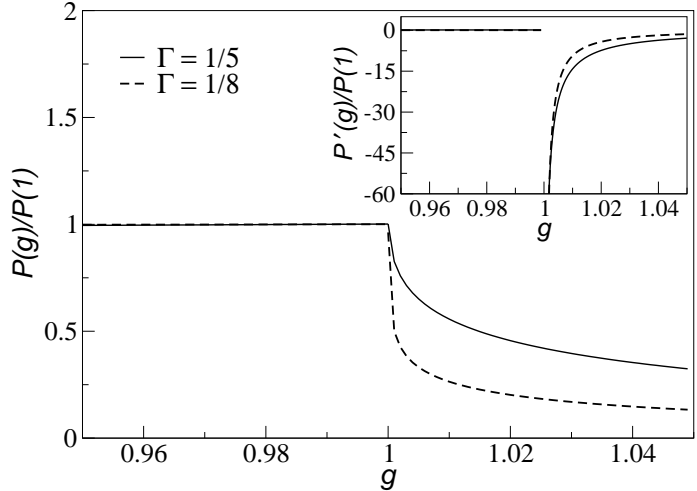}
\end{center}
\caption{$P(g)$ and its derivative $P'(g)$ (inset) around $g=1$ in
the insulating regime. The discontinuity in $P(g)$ is at
$g=1+\alpha$ \cite{mwgg}. Here $\alpha \sim 10^{-6}$ ($10^{-8}$)
for $\Gamma =1 /5$ ($\Gamma=1/8$).}
\end{figure}


\section{\label{betadif2}The case $\beta \ne 2$}

In this section we come back to our simpler one separated eigenvalue
approximation.  
As in section 2, we will use the  metallic limit Eq. (5) to be
true for all regimes. Then the shift approximation and the
corresponding expression for density  carries through, the only
difference is that there are three additional terms to be added to
the $V_{tot}$, defined in (24). One is obtained  from the last term in
(15),
while considering
the continuum approximation: rewriting this last term in the $\lambda$
variable, using $\lambda=\sinh^2 x$, and then doing the change of variable
$\sinh t = \lambda - \lambda_2$, as in Eq. (34), we have
\begin{equation}
 V^{\alpha}_0(t)=\alpha [\ln ({\rm arsinh}^2 \sqrt{\sinh^2
t+\lambda_2}) +\ln (\sinh^2 t+\lambda_2)+\ln (\cosh^2
t+\lambda_2)].
\end{equation}
The same last term in (15) also gives a direct term from the
separation of the lowest
level,
\begin{equation}
V^{\alpha}_1(x_1)=2\alpha \ln (x_1\sinh (2x_1)).
\end{equation}
A third term comes from the `entropy' term of Dyson \cite{dyson} derived
as the second term in (23)
\begin{equation}
V^{\alpha}_2(t)=-4\alpha \ln \sigma (t).
\end{equation}
It is normally assumed that this term does not change the density
appreciably, so the density appearing in (44) can be taken to be the
density due to the other terms. However, it includes $\sigma_z$,
and therefore the free energy is no longer quadratic in $z$.
This means that instead of doing the $\tau$ integral exactly
analytically, we have to do it either numerically, or try for a
saddle point solution. On the other hand, since the largest
density in the metallic regime is $\sigma_{\Gamma}$, we can expand
the logarithm and keep
only the leading term in $z$. As we will see, this will be sufficient
to give the  leading correction to the variance of $g$ in this regime.

\subsection{Correction to the metallic mean conductance}

The leading contribution to $F^{\alpha}_{z1}$ determines the leading
correction to the mean conductance $\langle g \rangle$.
We have
\begin{equation}
F^{\alpha}_{z1}=\frac{\beta}{2}\int dt V_z \sigma^{\alpha}_1,
\end{equation}
where the density is obtained from the potential
$V^{\alpha}_1(t)$, Eq. (105). Comparing with the potentials we
already have, we can rewrite it as
\begin{equation}
V^{\alpha}_1(t)=-2\alpha \left( V_{u_2}\vert_{x_1\rightarrow 0}
+V_{u_1}\vert_{\lambda_d\rightarrow \lambda_2}
+V_{u_1}\vert_{\lambda_d\rightarrow 1+\lambda_2} \right).
\end{equation}
Note that the last term is a $\lambda_d > 1$ term but will contribute
in the metallic limit when $\lambda_2 \ll 1$.
Thus the free energy can be written down in terms of known results
\begin{equation}
F^{\alpha}_{z1}=-2\alpha \left( F_{zu_2}\vert_{x_1\rightarrow 0}
+F_{zu_1}\vert_{\lambda_d\rightarrow \lambda_2}
+F_{zu_1}\vert_{\lambda_d\rightarrow 1+\lambda_2} \right).
\end{equation}
Collecting all the (already known) results, we obtain
in the metallic limit
\begin{equation}
F^{\alpha}_{z1} \approx \alpha z \left( \frac{4}{3}-b_2 x_2 \right).
\end{equation}
Since there is a constant term independent of $x_i$ and this
contributes to $F^{\prime}$, we immediately get a correction to the
average conductance $\Gamma$, given by
\begin{equation}
\langle g \rangle =\Gamma + \alpha \frac{8}{3}
=\Gamma +\left(1-\frac{2}{\beta}\right)\frac{1}{3},
\end{equation}
where we have included a factor of two for the symmetric term in the
free energy. The correction agrees with exact result.

\subsection{Correction to the metallic variance of conductance}

The leading correction to the variance comes from the contribution to
$F^{\alpha z}_{z2}$ from $V^{\alpha}_2(t)$, Eq. (107). Note that
$\sigma_{\Gamma}$ is the largest term, except very close to $t=0$
in a small range where $\sigma_{u_1}$ and $\sigma_{u_2}$ become
comparable. We neglect this complication, and in order to extract
the $z^2$ term, take as a first approximation $\sigma (t)\approx
\sigma_{\Gamma} (t)+\sigma_z (t)$. Then using $\sigma_{\Gamma} (t)
\gg \sigma_z (t)$, we expand the logarithm,
$V^{\alpha}_2(t)\approx \ln \sigma_{\Gamma}
+\sigma_z/\sigma_{\Gamma}$. The first term to leading
order will give a constant which we ignore, and a term
proportional to $x^2_2$ in the metallic limit. The second term to
leading order would consist of taking the $\lambda_2\rightarrow 0$
limit of both the densities. Then we have
\begin{equation}
F^{\alpha z}_{z2}=\frac{\beta}{2}\int dt
(-4\alpha)\frac{1}{\Gamma} \sigma_z; \;\;\; \sigma_z
=-\frac{8\alpha z^2}{\beta\Gamma}\int_0^{\infty}dt
\left[\frac{\pi^2/4-3t^2}{(t^2+\pi^2/4)^3}\right]^2.
\end{equation}
The result of this last integral is $12/\pi^3$. Then we have
\begin{equation}
F^{\alpha z}_{z2}=-\frac{1}{2}z^2
\frac{1}{\beta\Gamma}\left(1-\frac{2}{\beta}\right)\frac{24}{\pi^6},
\end{equation}
which gives a correction to the variance
\begin{equation}
\delta (var(g))=\frac{1}{\beta\Gamma}\left(1-\frac{2}{\beta}\right)
\frac{24}{\pi^6}.
\end{equation}
The exact result \cite{mirlin} can be written as $\delta
(var(g))=\frac{1}{\beta\Gamma}(1-\frac{2}{\beta})\mu$, with
$\mu=8/315=0.0254$, while for us, $\mu = 24/\pi^6=0.0250$, which
is very close. The $\beta$ and $\Gamma$ dependence are also
correct (in \cite{mirlin}, $\zeta/L=2\Gamma$; $\langle g \rangle$
and $var(g)$ have a factor 2 and 4 from spin, respectively).

\subsection{The density $\sigma^{\alpha}_1$}

Since the potential  $V^{\alpha}_1$ can be written in terms of potential
already considered, the corresponding density $\sigma^{\alpha}_1$ can also
be written down from the known densities, given as
\begin{equation}
\sigma^{\alpha}_1(t)=-2\alpha \left(\sigma_{u_2}\vert_{x_1\rightarrow 0}
+\sigma_{u_1}\vert_{\lambda_d\rightarrow \lambda_2}
+\sigma_{u_1}\vert_{\lambda_d\rightarrow 1+\lambda_2} \right) .
\end{equation}
In the metallic limit this simplifies to
\begin{equation}
\sigma^{\alpha}_1(t)\approx 8\alpha\left[\frac{1}{4t^2+\pi^2}
+\frac{1}{2\pi} \frac{x_2}{t^2+x^2_2}\right] ,
\end{equation}
while in the limit $x_2\rightarrow 0$,
\begin{equation}
\sigma^{\alpha}_1(t)\rightarrow  8\alpha\left[\frac{1}{4t^2+\pi^2}
+\frac{1}{4} \delta (t-0^+)\right] ,
\end{equation}
which agrees with \cite{beenakker1}.

\section{\label{summary}Summary and conclusion}

Given the distribution of the transmission eigenvalues, we have
developed a simple  method to obtain the full distribution of
conductances $P(g)$ at all strengths of disorder. The method is
based on the mapping of the $N$ eigenvalues to an electrostatic
problem of $N$ charges on the positive line with one- and two-body
interactions. Instead of treating all the charges as part of a
continuum in the $N\rightarrow\infty$ limit, we introduce the idea
of separating out a finite number of charges close to the origin
(corresponding to the most conducting channels) and treating the
rest as a continuum. This allows us to treat the strong disorder
region as well where the continuum model breaks down due to the
large separation between the charges, and where a finite number of
discrete charges give a better description. While the set of
discrete charges are treated exactly, the continuum part is
treated within a saddle point approximation to obtain a saddle
point free energy, from which all the relevant transport
quantities are evaluated. The saddle point density of the
continuum and its relation to the set of discrete charges show how
the nature of the solution changes qualitatively with increasing
disorder, and allows us to look for rare but allowed
configurations that give rise to novel characteristics of the
distribution. While two such characteristics, namely the existence
of non-analyticity near $g=1$ and the resulting asymmetry in the
distribution have been reported earlier, we show here for the
first time within the same framework that there exists a
log-normal tail in $P(g)$ for $g \ll 1$  even for weak disorder
where the average $\langle g \rangle \gg 1$, consistent with
earlier field theory calculations. In order to check the accuracy
of our method, we also obtain several other results in the
metallic limit for which exact results are available. These
include the average and the variance of conductance to order
$1/\Gamma$ for all symmetry classes and the difference in the
saddle point density between different symmetry classes.

The method developed here is very general and can be applied to
other quantities as well, {\it e.g.} the study of the distribution of
shot noise as well as the  conductance distribution in the
presence of Andreev scattering in an NS (Normal
metal-Superconductor) junction.

One should keep in mind that the DMPK approach does not contain
the effects of wave function correlations in the transverse
direction, which are expected to be important in higher dimensions
\cite{mg}. Nonetheless, the novel features in $P(g)$ near $g=1$
within our framework appears under very robust conditions which
are independent of dimensionality. It is therefore possible that
some of the features will persist in higher dimensions as well.
The similarities of the shape of $P(g)$ in the crossover regime
obtained within the current framework with the numerically
determined $P(g)$ in 3D at the critical point appears to be
consistent with the above conjecture.

This work has been supported in part by SFB 195  and by the Center for
Functional
Nanostructures of the
der Deutsche
Forschungsgemeinschaft. P.W. acknowledges support through a Max-Planck
Research Award.
V. A. G. is grateful for hospitality at the
IPCMS-GEMME, Strasbourg (France) where part of this work was done.

\appendix

\section{Partial densities}

In the following we give some details of the calculations for the
partial densities mentioned in Sec. 4.

It is convenient to define the integral:
\begin{equation}
\Omega_a(q)=\int_0^{\infty}ds \sin (qs)\frac{d}{ds}
V_a(\zeta(s)+\lambda_2).
\end{equation}
Then the partial density $\sigma_a $ is given by
\begin{equation}
\sigma_a=\frac{2}{\pi^2}\int_0^{\infty}dq \cos (qt)
(1-e^{-\pi q})\Omega_a(q).
\end{equation}
\begin{itemize}

\item Partial density $\sigma_z$:

Using the above definition with $V_z$ given by Eq. (46) we have
\begin{equation}
\Omega_z(q)=\int_0^{\infty}ds \sin (qs)\frac{d}{ds}
\frac{z}{\cosh^2 s+\lambda_2}
\end{equation}
which can be done by partial integration with result \cite{gradshteyn}:
\begin{equation}
\Omega_z(q)=-z \frac{\pi}{2}
\frac{1}{\sqrt{\lambda_2(1+\lambda_2)}}
\frac{q \sin (q a/2)}{\sinh (q \pi/2)}
\end{equation}
where
\begin{equation}
a={\rm arcosh}(1+2\lambda_2)=2{\rm arsinh} \sqrt{\lambda_2}=2x_2.
\end{equation}
Therefore the density $\sigma_z$ can be written as
\begin{equation}
\sigma_z(t)=\frac{2}{\pi}
\frac{z}{\sqrt{\lambda_2(1+\lambda_2)}}
\frac{\partial}{\partial s}
\frac{\partial}{\partial \gamma}
\int_0^{\infty}dq \frac{\sin (qs)}{q}\sin \left(\frac{q 2x_2}{2}\right)
e^{-\gamma q}
\end{equation}
evaluated at $\gamma=\pi /2$. From \cite{gradshteyn} and after
algebraic simplification we obtain $\sigma_z$ as given in (59a).


\item Partial density $\sigma_{u1}$

We will first consider the metallic regime. Taking the derivative
of $V_{u1}$ we have
\begin{equation}
\Omega_{u_1}=-\frac{1}{2}\int_0^{\infty}ds
\frac{\sin (qs) \sinh (2s)}{\sinh^2 s+ \lambda_2-\lambda_1}
\end{equation}
and using $\sinh^2 s=(\cosh (2s)-1)/2$ then $\Omega_{u_1}$ is given by
\begin{equation}
\Omega_{u_1}=-\frac{\pi}{2}
\frac{\cosh (qa_0/2)}{\sinh (q\pi/2)}, \;\;\;{\rm for} \; \lambda_d < 1,
\end{equation}
where $\lambda_d=\lambda_2-\lambda_1 $ and
$a_0=\pi-2\arcsin\sqrt{\lambda_d}$. Finally, after writing
$(1-e^{-\pi q})=2e^{-\pi q/2}\sinh (\pi q/2)$ in (A2), $\sigma_{u1}$
can be written as given in (59b) for $\lambda_d < 1$.

In order to obtain $\sigma_{u1}$ for $\lambda_d>1$, we follow a
different method. First we take the derivative w.r.t. $\lambda_d$:
\begin{equation}
\frac{\partial \Omega_{u_1}}{\partial \lambda_d}=
-\frac{1}{2}\int_0^{\infty}ds
\sin (qs) \frac{d}{ds}\frac{1}{\sinh^2 s+ \lambda_d}.
\end{equation}
A partial integration gives the boundary term zero and
\begin{equation}
\frac{\partial \Omega_{u_1}}{\partial \lambda_d}=\frac{q}{2}
\frac{\partial I_{\Omega_1}}{\partial \lambda_d}
\end{equation}
where
\begin{equation}
\frac{\partial I_{\Omega_1}}{\partial \lambda_d}
=2\int_0^{\infty}ds\frac{\cos (qs)}{\cosh (2s)+(2 \lambda_d-1)}.
\end{equation}
For $\lambda_d >1$ \cite{gradshteyn},
\begin{equation}
\frac{\partial I^>_{\Omega_1}}{\partial \lambda_d}
=\frac{\pi}{\sinh (q\pi/2)}
\frac{\sin [(q/2){\rm arcosh} (2\lambda_d-1)]}{\sqrt{(2\lambda_d-1)^2-1}}.
\end{equation}
Using ${\rm arcosh} (2\lambda_d-1)=y$, we can integrate to obtain
\begin{equation}
I^>_{\Omega_1}=-\frac{\pi}{q}\frac{\cos (qy/2)}{\sinh (q\pi/2)}
+\textrm{terms independent of } \lambda_d.
\end{equation}
Neglecting the $\lambda_d$-independent terms,
then we have a result valid for $\lambda_d >1$. Therefore we have
\begin{equation}
\Omega_{u_1}=-\frac{\pi}{2\sinh (q\pi /2)}\left\{ \begin{array}{ll}
\cosh (qa_0 /2), &  \lambda_d < 1; \\
\cos (qb_0 /2), & \lambda_d > 1;\end{array} \right.
\end{equation}
where we have included the result for $\lambda_d <1$ obtained within
this method (see Eq. (A8)) and $b_0={\rm arcosh}(2\lambda_d-1)$. We
can check that there is no $\lambda_d$ independent terms by
comparing the two expressions of $\Omega_{u_1}$ at the boundary
$\lambda_d=1$ {\it i.e.} at $a_0=0$ and $b_0=0$ for $\lambda_d <1$
and $\lambda_d > 1$, respectively. We note that $b_0=ia_0$, and
the expression for $\lambda_d < 1$ becomes valid for $\lambda_d
> 1$ if we allow $a_0$ to be complex. The corresponding $\sigma_{u_1}$
for all $\lambda_d$ is then given by (59b).

\item Partial density $\sigma_{u2}$

Here we will use our approximation to $V'_{u2}(s)$, Eq. (58). This
approximation starts to fail for $x_2 >5$  and/or $x_d \ll 1$,
however the metallic limit $x_2 \to 0$ is well described and also
it has been shown in \cite{gmw} that this approximation remains
valid in the crossover regime. Equation (58) can also be written
as
\begin{equation}
V'_{u_2}(s)\approx\frac{s}{2}\left[\frac{1}{(s-ic)^2+\beta^2_1}
+\frac{1}{(s+ic)^2+\beta^2_1}\right]-\frac{a s^2_m s
}{(s^2+s^2_m)^2+4c^2s^2}
\end{equation}
with $\beta_1^2=c^2+s_m^2$. Introducing this expression in (A1), we get
\begin{equation}
\Omega_{u_2}(q)\approx -\frac{\partial}{\partial q}
\int_0^{\infty}\frac{ds}{s} \cos (qs) V'_{u_2}(s)
=-\frac{\pi}{4\beta_1}\left[A_+e^{-q(\beta_1+c)}+A_-
e^{-q(\beta_1-c)}\right]
\end{equation}
where $A_{\pm}=\beta_1\pm c \pm as^2_m/2c$, and the corresponding
density becomes as given in (59c).

\item Partial density $\sigma_{\Gamma}$

In order to obtain $\sigma_{\Gamma}$ we will use our approximation to
$V'_{\Gamma}(s)$, Eq. (57). In this case, we have
\begin{eqnarray}
\Omega_{\Gamma}&\approx&\Gamma\int_0^{\infty}ds \sin (qs) s
\left[1-\frac{a_4}{\cosh(2s)+\cosh\delta_1}\right] \nonumber \\
&=& -\pi\delta'(q)+a_4
\frac{\partial}{\partial q}\int_0^{\infty}ds
\frac{\cos (qs)}{\cosh( 2s)+\cosh\delta_1} \nonumber \\
&=&-\pi\delta'(q)+a_4 \frac{\pi}{2\sinh\delta_1}\frac{\partial}{\partial q}
\frac{\sin (q\delta_1/2)}{\sinh (q\pi /2)}
\end{eqnarray}
Therefore $\sigma_\Gamma$ is given by
\begin{equation}
\sigma_\Gamma(t)\approx \frac{2}{\pi^2}
\int_0^{\infty}dq \cos qt (1-e^{-\pi q})\left[-\pi\delta'(q)
+\frac{\pi a_4}{2\sinh\delta_1}\frac{\partial}{\partial q}
\frac{\sin (q\delta_1/2)}{\sinh (q\pi /2)}\right].
\end{equation}
The integrals for each term in (A22) can be done. The final result
is given in (59d).

\item Partial density $\sigma_2$

We first need to evaluate the potential $V_2$ given by (33), which
can now be done with the densities evaluated above. Analytic
expressions are possible only in the metallic regime, where the
largest contributions come from $\sigma_{\Gamma}$ (which is
proportional to $\Gamma \gg 1$) and $\sigma_z$. Since $V_2$ itself
is proportional to $\lambda_2$, we will ignore the contributions
from $\sigma_{u_1}$ and $\sigma_{u_2}$ as well as from $\sigma_2$,
which will otherwise require a self consistent calculation.

The potential $V_{2\Gamma}$ obtained from $\sigma_{\Gamma}$ is
\begin{equation}
V_{2\Gamma}=-\Gamma\lambda_2 A(t)
\end{equation}
where we have defined
\begin{equation}
A(t)=\int_0^{\infty}ds\frac{1}{t^2-s^2} \left[\frac{t}{\sinh
(2t)}-\frac{s}{\sinh (2s)}\right].
\end{equation}
and we have neglected the $\lambda_2$ dependent part of
$\sigma_{\Gamma}$. We extend the integral to $-\infty$ and
continue to the complex upper half plane. Note that there is no
pole on the real line. Using variable $u=2s$ and keeping in mind
that poles on the real line are to be excluded, we can rewrite it
as
\begin{equation}
A=-\frac{1}{2}\int_c du\frac{1}{4t^2-u^2} \frac{u}{\sinh (u)}.
\end{equation}
The poles are at $u=ik\pi$ for $k=$integers. This gives
\begin{equation}
A=\sum_{k=1}^{\infty}\frac{k\cos (k\pi)} {k^2+4t^2/\pi^2}.
\end{equation}
The corresponding $\Omega_{2\Gamma}$, after a partial integration,
can be written as
\begin{equation}
\Omega_{2\Gamma}=\Gamma\lambda_2 \frac{\pi^2}{8}q \left[\tanh
(q\pi/4)-1\right].
\end{equation}
The density for this part is, after algebraic simplification,
\begin{equation}
\sigma_{2\Gamma}=\frac{\Gamma\lambda_2}{2}\left[
\frac{t^2-\pi^2/4}{(t^2+\pi^2/4)^2}
-\frac{t^2-\pi^2}{(t^2+\pi^2)^2}\right].
\end{equation}

\section{ Partial free energies}

With the above partial densities of Appendix A, it is possible in
principle calculate all the partial free energies $F_{ab}$ through
the relation (48). However, only $F_{zz}$, $F_{zu_1}$ and
$F_{u_1u_1}$ are evaluated analytically exactly. For some of the
remaining free energy terms we made some approximations
which gives both the exact metallic and insulating
regimes and a reasonable approximation (as checked numerically) in
the intermediate regime, while other free energies require
different sets of approximations in different regimes. In the
following we will point out the approximations used to obtain each
expression in the paper.

\item Free energy $F_{zz}$

This can be obtained exactly. Although we have evaluated
$\sigma_z$ already, it is useful to start with the full expression
of $F_{zz}$ in terms of the three integrals (Eqs. 48, 49 ) because
as we will see, changing the order of integrations  (e.g. doing
the $q$ integral last) will sometimes be more convenient. Using
the result for $\Omega_z$ in (A4), writing $\cosh^2 t=(\cosh
(2t)+1)/2$, and $1-e^{-\pi q}=2e^{-\pi q/2}\sinh (\pi q/2)$, we
have
\begin{equation}
F_{zz}=-\frac{2z^2}{\pi\beta}
\frac{1}{\sqrt{\lambda_2(1+\lambda_2)}} \int_0^{\infty}dq q\sin
(qx_2) e^{-\pi q/2} \int_0^{\infty}dt\frac{\cos (qt)}{\cosh
(2t)+1+2\lambda_2}
\end{equation}
The $t$ integral can be done. Changing variables $x=\pi q/2$ and
writing $\sin^2 (\theta)=(1-\cos (2\theta))/2$, the result is
\begin{equation}
F_{zz}=-\frac{z^2}{2\pi\beta}\frac{1}{\lambda_2(1+\lambda_2)}
\int_0^{\infty}dx \frac{2x}{\pi} \left[1-\cos
\left(\frac{2a}{\pi}x\right)\right] \frac{e^{-x}}{\sinh x}.
\end{equation}
where $a=2x_2$ has been defined for later convenience. The first
term is known. The second integral can also be done by noting that
$(2x/\pi)\cos (2a\pi x)$ can be rewritten as $(\partial/\partial
a)\sin (2a\pi x)$. The final result is given by (66), valid for all
$\lambda_2$. This is a crucial term that dictates the crossover
from metallic to insulating behavior. Note that the leading term
for $\lambda_2 \ll 1$, after cancellations, is
\begin{equation}
F_{zz}=-\frac{z^2}{15\beta}\left[1+O(\lambda_2)\right].
\end{equation}
For $\lambda_2=0$, this gives the exact variance in the zeroth
order.

\item Free energy $F_{\Gamma z}$

Using the expression for $\sigma_z$ (Eq. 59a), we obtain by
partial integration:
\begin{eqnarray}
F_{\Gamma z}=\frac{\Gamma zx_2}{2\sqrt{\lambda_2(1+\lambda_2)}}
\int_0^{\infty}dt &&\frac{{\rm arsinh} \sqrt{\sinh^2 t+\lambda_2}}
{\sqrt{(\cosh^2 t+\lambda_2)(\sinh^2 t+\lambda_2)}} \times \nonumber \\
&&\frac{t\sinh (2t)}{(t^2+(\pi^2+a^2)/4)^2-a^2t^2}.
\end{eqnarray}

We first obtain an expression valid in the metallic limit.

Expand the
integrand in powers of $\lambda_2$ up to the first power. The
leading $\lambda_2$-independent term leads to a simple integral,
giving
\begin{equation}
F^{1(met)}_{\Gamma z}=\frac{z\Gamma
x_2}{2\sqrt{\lambda_2(1+\lambda_2)}}=\frac{z\Gamma x_2}{\sinh
(2x_2)}.
\end{equation}
The next term involves
\begin{equation}
I_{\Gamma z}=\int_0^{\infty}dt
\left[\frac{1}{(t^2+\pi^2/4)^2}\frac{t}{\sinh 2t}
-\frac{t^2}{2(t^2+\pi^2/4)^2}\left(\frac{1}{\sinh^2 t}\right)
\right].
\end{equation}
where a term $2(t^2-\pi^2/4)/(t^2+\pi^2/4)^4$ in the integrand
coming from differentiating the $a=2x_2$ dependent factor in (B4)
has been omitted because its integral over $t$ is zero. The first term in
(B6)
is of the form
\begin{eqnarray}
I_1&=&-\frac{\partial}{\partial b}\int_0^{\infty}\frac{dt}{t^2+b}
\frac{t}{\sinh (2t)} =-\frac{\partial}{\partial
b}\left[\frac{\pi}{4\sqrt{b}}
+\pi\sum_{k=1}^{\infty}\frac{(-1)^k}{2\sqrt{b}+k\pi}\right]. \\
  &=&\frac{1}{\pi^2}\left[1+2\sum_{k=1}^{\infty}\frac{(-1)^k}
{(k+1)^2}\right] = \frac{1}{\pi^2}[\zeta(2)-1)].
\end{eqnarray}
Writing $\frac{1}{\sinh^2 t}+\frac{1}{\cosh^2 t}
=-2\frac{d}{dt}\frac{1}{\sinh (2t)}$, doing a partial integration
and then using
$\frac{t^3}{(t^2+\pi^2/4)^3}=\frac{t}{(t^2+\pi^2/4)^2}
[1-\frac{\pi^2/4}{(t^2+\pi^2/4)}]$, the second term in (B6) can be
rewritten as
\begin{eqnarray}
I_2&=&-\int_0^{\infty}\frac{dt}{\sinh (2t)}\left[
-\frac{2t}{(t^2+\pi^2/4)^2}+\frac{\pi^2 t}{(t^2+\pi^2/4)^3}
\right]. \\
&=&-\frac{1}{\pi^2}[3\zeta(3)-\zeta(2)-1],
\end{eqnarray}
where $\zeta(x)$ is the Zeta function. Collecting terms, the free
energy is
\begin{equation}
F^{(met)}_{\Gamma z}\approx -\frac{z\Gamma x^2_2}{3}
-\frac{z\Gamma x^2_2}{\pi^2}[3\zeta(3)-2\zeta(2)],
\end{equation}
where we have used $\lambda_2\rightarrow 0$ limit and kept only
the leading $x_2$ dependent terms.

To obtain an expression valid in the crossover regime, we use the
fact that $\sigma_z$ can be integrated exactly. We define
\begin{equation}
\omega_z=\int dt \sigma_z (t)=\frac{z}{2\sinh 2x_2}
\left[\frac{1}{(t+x_2)^2+\pi^2/4}-\frac{1}{(t-x_2)^2+\pi^2/4}\right].
\end{equation}
Then
\begin{equation}
F_{\Gamma z}=\int_0^{\infty} dt V_{\Gamma}\sigma_z
=V_{\Gamma}\omega_z|^{\infty}_0-\int_0^{\infty} dt
V'_{\Gamma}\omega_z.
\end{equation}
The first term is zero, and we use the approximation (57) for
$V'_{\Gamma}$ to obtain, after simple integrals,
\begin{equation}
F^{(cross)}_{\Gamma z}=\frac{z\Gamma x_2}{\sinh
2x_2}(1+\frac{a_4}{2 x_2}B),
\end{equation}
where
\begin{equation}
B=\int_{-\infty}^{\infty} dz \frac{z}{\cosh z +\cosh\delta_1}
\left[
\frac{1}{(z+2x_2)^2+\pi^2}-\frac{1}{(z-2x_2)^2+\pi^2}\right].
\end{equation}
This can be done by contour integration. The integrand has poles
at
\begin{equation}
z=\pm i\pi \pm 2x_2; \;\;\; z=\pm i(2k+1)\pi \pm \delta_1, \;\;\;
k=0,1,2,\cdots
\end{equation}
Closing the contour on the upper half plane, the sum over residues
can be rearranged and identified with DiGamma functions $\Psi(x)$:
\begin{eqnarray}
B&&=\frac{2x_2}{\cosh 2x_2-\cosh\delta_1}
-\frac{4x_2}{4x^2_2-\delta^2_1}\frac{\delta_1}{\sinh\delta_1} \nonumber \\
&-&\frac{1}{2\sinh\delta_1}
\left(\Psi[1+ix_{2-}]-\Psi[1-ix_{2+}]-\Psi[1+ix_{2+}]
+\Psi[1-ix_{2-}]\right), \nonumber \\
\end{eqnarray}
where $x_{2\pm}=x_2\pm \delta_1/2$.

\item Free energy $F_{zu_1}$

This is done exactly.  Using the expression for $V_z$ in (46) and
$\Omega_{u_1}$ in (A14), we get
\begin{equation}
F_{zu_1}=-\frac{z}{\pi}\int_0^{\infty}\frac{dt}{\cosh^2
t+\lambda_2} \int_0^{\infty}dq \cos (qt)\cosh (qa_0/2)e^{-q\pi
/2}.
\end{equation}
We do the t-integral first, writing $\cosh^2 t$ in terms of $\cosh
(2t)$. For the $q$-integral,  we use $e^{-q\pi /2}/\sinh (q\pi
/2)=2/(e^{q\pi}-1)$, giving
\begin{equation}
F_{zu_1}=\frac{z}{\sqrt{\lambda_2(1+\lambda_2)}}
\left[\frac{a}{a^2+a_0^2}-\frac{1}{2}\frac{\sinh a} {\cosh a-\cos
a_0}\right].
\end{equation}
Finally, using $\lambda_i=\sinh^2 x_i$, $a=2x_2$, and $\cosh
a-\cos a_0=2(1+\lambda_1)$, we get
\begin{equation}
F_{zu_1}=\frac{z}{\sinh(2x_2)}
\left[\frac{4x_2}{4x^2_2+a_0^2}-\frac{\sinh(2x_2)}{2\cosh^2 x_1
}\right].
\end{equation}
As mentioned after (A14), $a_0$ is imaginary for $\lambda_d > 1$.
Using $\Omega_{u_1}$ for $\lambda_d > 1$ from (A14) gives the same
result.

The metallic limit of (B20) is simply
\begin{equation}
F^{(met)}_{zu_1}\approx \frac{z8x_d}{\pi^3}.
\end{equation}

\item Free energy $F_{zu_2}$:

Using the expression for $V_z$ and $\Omega_{u_2}$, writing
$\cosh^2 t$ in terms of $\cosh (2t)$ and doing the $t$-integral
first as in $F_{zu_1}$ results in elementary integrals over $q$.
The result is
\begin{equation}
F_{zu_2}\approx -\frac{z}{4\beta_1}\frac{2x_2}{\sinh (2x_2)}\left[
\frac{A_+}{x^2_2+(\pi/2+\beta_1+c)^2}+(c\rightarrow -c)\right].
\end{equation}
Note that $\Omega_{u_2}$ itself uses an approximate expression for
$V_{u_2}$. The free energy expression is a good approximation in
both the metallic and the crossover regimes as checked
numerically. In the metallic limit, $A_{\pm}\rightarrow
\beta_1\rightarrow x_d$, $c\rightarrow 0$ and (B22) gives
\begin{equation}
F^{(met)}_{zu_2}\approx \frac{z8x_d}{\pi^3}.
\end{equation}

\item Free energy $F_{z2}$:

Since $\sigma_2$ is valid only in the metallic limit, $F_{z2}$ can
be evaluated only in the metallic limit. We use the
$\lambda_2\rightarrow 0$ limit of $V_z$ to get
\begin{equation}
F^{(met)}_{z2}\approx\frac{z\Gamma\lambda_2}{2}\sum_{k=1}^{\infty}\cos
(k\pi) \int_0^{\infty}\frac{dt}{\cosh^2 t} \int_0^{\infty}dq q\cos
(qt)\sinh (\pi q/2) e^{-\frac{\pi q}{2}(1+k)}.
\end{equation}
Doing the $t$-integral first and then the $q$ integral, we get
\begin{equation}
F_{z2}\approx\frac{z\Gamma\lambda_24}{\pi^2}\sum_{k=1}^{\infty}
\frac{\cos (k\pi)}{(k+1)^3}.
\end{equation}
The sum can be rewritten as
\begin{equation}
\sum_{k=1}^{\infty}\frac{\cos (k\pi)}{(k+1)^3}
=-\sum_{k=2}^{\infty}\frac{(-1)^k}{k^3}
=-\left[-\sum_{k=1}^{\infty}\frac{(-1)^{k+1}}{k^3}+1\right],
\end{equation}
and evaluated in terms of Zeta functions, giving finally, for
$\lambda_2\ll 1$:
\begin{equation}
F^{(met)}_{z2}\approx\frac{z\Gamma\lambda_2}{\pi^2}[3\zeta (3)-4],
\end{equation}

\item Free energy $F_{\Gamma\Gamma}$:

This requires one approximation for the metallic limit, a second
one for the insulating limit, and a third one for the crossover
regime.

For the metallic limit, we use expansions of $V_{\Gamma}$ and
$\sigma_{\Gamma}$ in powers of $\lambda_2$:
\begin{equation}
F^{(met)}_{\Gamma\Gamma}\approx\frac{\beta\Gamma^2}{4}\int_0^{\infty}
dt
\left[t^2+\lambda_2\frac{2t}{\sinh(2t)}\right]\left[1-\frac{\lambda_2}{2}
\frac{\partial}{\partial t}\left( \frac{1}{t}-\frac{2}{\sinh
(2t)}-\frac{t^2}{t^2+\frac{\pi^2}{4}}\right)\right].
\end{equation}
The first term diverges. However, it is independent of $\lambda_2$
and so we ignore it. Rest of the integrals, keeping only terms up
to order $\lambda_2$, gives
\begin{equation}
F^{(met)}_{\Gamma\Gamma}\approx\frac{\beta\Gamma^2x^2_2\pi^2}{16},
\end{equation}
where we have replaced $\lambda_2 \ll 1$ by $x^2_2$.

In the insulating limit, we approximate $V'_{\Gamma}$ as
\begin{equation}
V'_{\Gamma}(s) \approx \frac{\Gamma}{2}\left\{
\begin{array}{ll}\frac{\sinh(2s) x_2}{\sinh^2 x_2},  & {\rm for} \;
s \ll x_2; \cr \frac{\sinh(2s)s}{\sinh^2 s}, & {\rm for} \; s \gg
x_2 ,
\end{array} \right. \nonumber \\
\end{equation}
Then we obtain $\sigma_{\Gamma}(t)$ for $t\ll x_2$ and also for
$t\gg x_2$ by dividing up the integral over $s$ from zero to $x_2$
and then from $x_2$ to infinity, using the approximation for
$V'_{\Gamma}$. We then obtain $F_{\Gamma\Gamma}$ by dividing the
integral over $t$ into one from zero to $x_2$ and the other from
$x_2$ to infinity. The final result is
\begin{equation}
F^{(ins)}_{\Gamma\Gamma}=\beta\Gamma^2x^2_2.
\end{equation}

In the crossover regime, we use $\sigma_{\Gamma}(t)$ from (59d),  
and $V_{\Gamma}(t)=V^0_{\Gamma}(t)+\delta V_{\Gamma}(t)$, where
$V^0_{\Gamma}=V_{\Gamma}(x_2\rightarrow 0)=\Gamma t^2/2$. The
first term in integral (48) coming from the first term of
$\sigma_{\Gamma}$ in (59e) involves
\begin{equation}
K^{1}_{\Gamma\Gamma}=\int_0^{\infty} dt V_{\Gamma}(t)
=\int_0^{\infty} dt \frac{1}{2}\Gamma t^2 + \int_0^{\infty} dt
\delta V_{\Gamma}(t).
\end{equation}
The first term diverges; but it is independent of $x_i$ and so we
ignore it. We do a partial integration for the second term and use
the approximation (57) for $\delta V'_{\Gamma}$, leading to
\begin{equation}
K^{1}_{\Gamma\Gamma}=a_4\Gamma\int_0^{\infty} dt \frac{t^2}{\cosh
2t+\cosh\delta_1} =-a_4\Gamma \left[\frac{1}{\lambda}
\frac{\partial}{\partial\lambda}\int_0^{\infty}dt \frac{\cos
\lambda t}{\cosh 2t+\cosh\delta_1}\right]_{\lambda=0}.
\end{equation}
Expanding the result for the integral in power series in $\lambda$
and taking the limit we obtain
\begin{equation}
K^{1}_{\Gamma\Gamma}=\frac{a_4\Gamma}{24}\frac{\delta_1}{\sinh\delta_1}
(\pi^2+\delta^2_1).
\end{equation}
The second term of $\sigma_{\Gamma}$ leads to integrals of the
form
\begin{equation}
K^{2}_{\Gamma\Gamma}=\int_0^{\infty} dt V_{\Gamma}(t)
\left[\frac{1}{t^2+c^2_+}-\frac{1}{t^2+c^2_-}\right].
\end{equation}
The dominant term $V^0_{\Gamma}(t)$ gives $\Gamma\pi(c_--c_+)/4$.
The next term $\delta V_\Gamma$ is again done by partial
integration, leading to
\begin{equation}
\Gamma a_4\int_0^{\infty} dt \frac{t}{\cosh 2t+\cosh\delta_1}
\left[\frac{1}{c_+}\arctan\frac{t}{c_+}-\frac{1}{c_-}\arctan\frac{t}{c_-}
\right].
\end{equation}
The third term of $\sigma_{\Gamma}$ leads to
\begin{equation}
K^{3}_{\Gamma\Gamma}=\int_0^{\infty} dt
V_{\Gamma}(t)\frac{1}{\cosh 2t+\cosh\delta_1}.
\end{equation}
Again the dominant term of  $V_{\Gamma}(t)$ gives
\begin{equation}
\frac{\Gamma} {48}\frac{\delta_1}{\sinh
\delta_1}(\pi^2+\delta^2_1).
\end{equation}
The next term, after a partial integral, leads to
\begin{equation}
\frac{\Gamma a_4}{\sinh\delta_1}\int_0^{\infty} dt \frac{t}{\cosh
2t+\cosh\delta_1} \tanh^{-1}\left(\tanh t \tanh\frac{\delta_1}{2}\right).
\end{equation}
Collecting all the terms with their appropriate pre-factors, we
get the expression
\begin{equation}
F_{\Gamma\Gamma}=\frac{\Gamma^2 a_4}{12}\frac{\delta_1}{\sinh
\delta_1}(\pi^2+\delta^2_1) +\frac{\Gamma^2
a^2_4}{i\pi\sinh\delta_1}\int_0^{\infty} dt \frac{t\Phi(t)}{\cosh
2t+\cosh\delta_1},
\end{equation}
where
\begin{eqnarray}
\Phi(t)=\frac{\pi+i\delta_1}{2}\arctan \frac{2t}{\pi-i\delta_1}
-\frac{\pi-i\delta_1}{2}\arctan \frac{2t}{\pi+i\delta_1}\nonumber \\
 -\pi \tanh^{-1}[\tanh t\tanh (\delta_1/2)]
\end{eqnarray}
The integral with $\Phi(t)$ can not be done. However, to a good
approximation, we can write
\begin{equation}
\Phi(t)\approx \frac{D_1 t}{(1+\sqrt{D_1/D_2}t^2)^2}; \;\;
D_1=\frac{4i\pi\delta_1}{\pi^2+\delta^2_1}-i\pi\tanh \frac{\delta_1}{2};
\;\; D_2=-i\frac{\pi\delta_1}{12}(\pi^2+\delta^2_1).
\end{equation}
Denoting $\eta'=\sqrt{D_1/D_2}$, we need the integral
\begin{equation}
\int_0^{\infty} dt \frac{1}{\cosh 2t+\cosh\delta_1}\frac{D_1
t^2}{(1+\eta't^2)^2} =-\frac{D_1}{2}\frac{\partial}{\partial\eta'}
\int_{-\infty}^{\infty} dt \frac{1}{\cosh
2t+\cosh\delta_1}\frac{1}{1+\eta't^2}.
\end{equation}
This integral is similar to (B15), and the result is
\begin{eqnarray}
\int_{-\infty}^{\infty} \frac{dt}{\cosh
2t+\cosh\delta_1}&&\frac{1}{t^2+a^2}
=\frac{1}{ia\sinh\delta_1}\nonumber \\ &&\times\left[\Psi\left(\frac{1}{2}
+\frac{a}{\pi}+ \frac{i\delta_1}{2\pi}\right)
-\Psi\left(\frac{1}{2}+\frac{a}{\pi}-\frac{i\delta_1}{2\pi}\right)\right].
\end{eqnarray}
Note that the integral is true for $\delta_1$ either real or
imaginary. Collecting terms, we finally have
\begin{eqnarray}
F_{\Gamma\Gamma}&=&\frac{\Gamma^2 a_4}{12}\frac{\delta_1}{\sinh
\delta_1}(\pi^2+\delta^2_1) \nonumber \\
&\times&\left[1-\frac{a_4}{i4\sinh\delta_1}\left(\frac{1}{\eta}
(\Psi[\eta_+]-\Psi[\eta_-])
+\frac{1}{\pi}(\Psi'[\eta_+]-\Psi'[\eta_-])\right)\right],
\end{eqnarray}
where
\begin{equation}
\eta_{\pm}=\frac{1}{2}+\frac{\eta\pm i\delta_1/2}{\pi},
\end{equation}
and  $\Psi'$ is the derivative of the DiGamma function.

\item Free energy $F_{\Gamma u_1}$:

This energy is analytically calculated in the metallic and
insulating regimes and numerically in the crossover region.

In the metallic limit, we use $\sigma_{u_1}$ and the
$\lambda_2\rightarrow 0$ limit of $V_{\Gamma}$:
\begin{equation}
F^{(met)}_{\Gamma u_1}\approx
-\frac{\beta\Gamma}{4\pi}\int_0^{\infty}dt t^2
\left[\frac{\mu_-}{t^2+\mu^2_-}+\frac{\mu_+}{t^2+\mu^2_+}\right];\;\;\;
\mu_{\pm}=\frac{\pi\pm a_0}{2}.
\end{equation}
Rewriting $
\mu_i
t^2/(t^2+\mu^2_i)=\mu_i\left(1-\mu^2_i/(t^2+\mu^2_i)\right)$
and the relation $\mu_-+\mu_+=\pi$, we get
\begin{equation}
F^{(met)}_{\Gamma u_1}\approx
-\frac{\beta\Gamma}{4\pi}\int_0^{\infty}dt
\left[\pi-\left(\frac{\mu^3_-}{t^2+\mu^2_-}+\frac{\mu^3_+}{t^2+\mu^2_+}\right)
\right].
\end{equation}
The first term is divergent but constant, so we ignore it. The
second term, after using the definitions of $\mu_i$, gives
\begin{equation}
F^{(met)}_{\Gamma u_1}\approx\frac{\beta\Gamma}{16}a^2_0.
\end{equation}
Using $a_0=\pi-2\arcsin\sqrt{\lambda_d}$ and $\lambda_d \ll 1$,
and ignoring constants we finally obtain
\begin{equation}
F^{(met)}_{\Gamma u_1}\approx -\frac{\beta\Gamma}{4}\pi x_d.
\end{equation}

In the insulating limit, $\sigma^{(ins)}_{\Gamma}\approx 2\Gamma/
x_2$ which gives
\begin{equation}
F^{(ins)}_{\Gamma u_1}\approx -\frac{\beta\Gamma}{2}3x_2.
\end{equation}

\item Free energy $F_{\Gamma u_2}$:

$F_{\Gamma u_2}$ is calculated only in the metallic
limit while for the crossover region, it is obtained numerically.
Contribution in the insulating limit is negligible compared to
$F^{(ins)}_{\Gamma u_1}$, and will not be evaluated.

In the metallic limit we proceed as for $F_{\Gamma u_1}$. Using
$\sigma_{u_2}$, the dominant term $V^0_{\Gamma}(t)=\Gamma t^2/2$
(defined above (B31)) gives a diverging constant that we ignore,
plus a term
\begin{equation}
F^{(met)}_{\Gamma u_2}\approx-\frac{\beta\Gamma\pi\beta_1}{4}
\end{equation}
where we have kept only the leading term in powers of $\lambda_2
\ll 1 $. Since in this limit $\beta_1\rightarrow x_d$, we obtain
\begin{equation}
F^{(met)}_{\Gamma u_2}\approx-\frac{\beta\Gamma}{4}\pi x_d.
\end{equation}

\item Free energy $F_{u_1u_1}$:

This can be done exactly. We use $V_{u_1}$ from (46) and
$\Omega_{u_1}$ from (A14). We do the $t$ integral first,
\begin{equation}
I_{u_1 u_1}=\int_0^{\infty}dt \ln (\sinh^2 t+\lambda_d)\cos (qt).
\end{equation}
A partial integration gives a diverging term. We introduce a cut
off $b$:
\begin{equation}
I_{u_1 u_1}=\frac{2}{q}[b\sin (qb)]_{b\rightarrow \infty}
-\frac{\pi}{q}\frac{\cosh (qa_0 /2)}{\sinh (q\pi /2)}
\end{equation}
The $q$-integral for the free energy leads to a diverging constant
from the first term and we ignore it. The second term can be
rewritten in the form
\begin{equation}
F_{u_1 u_1}=-\frac{\beta}{2} \int_0^{\infty}\frac{dq}{q} e^{-\pi
q/2}\frac{1+\sinh^2 (qa_0/2)}{\sinh(q\pi /2)}.
\end{equation}
Using $x=q\pi/2$, this integral can be done, and we get
\begin{equation}
F_{u_1 u_1}=-\frac{\beta}{2}\left[\frac{1}{2}\ln \left(
\frac{a_0}{\sin a_0}\right)
+\int_0^{\infty}\frac{dx}{x}\frac{e^{-x}}{\sinh x}\right].
\end{equation}
Neglecting the diverging constant, we finally have
\begin{equation}
F_{u_1 u_1}=-\frac{\beta}{4}\ln \left(\frac{a_0}{\sin a_0}\right).
\end{equation}
Again, $a_0$ is imaginary for $\lambda_d > 1$.

\item Free energy $F_{u_2u_2}$:

Given the form (59c) for $\sigma_{u_2}$, this can be done exactly.
However, $\sigma_{u_2}$ itself has approximations. The resultant
approximate $F_{u_2u_2}$ will be valid in the crossover regime as
checked numerically, and will also give the correct metallic
limit.

Using $\sigma_{u_2}$ and a partial integration, we start with
\begin{eqnarray}
&&F_{u_2 u_2}\approx\frac{\beta}{4\pi\beta_1}\int_0^{\infty}dt
V'_{u_2}\times \nonumber \\ &&\left[
A_{+}\left(\arctan\frac{t}{\Lambda_1}-\arctan\frac{t}{\Lambda_2} \right) 
+A_{-}\left(\arctan\frac{t}{\Lambda_3}-\arctan\frac{t}{\Lambda_4}\right)
\right],
\end{eqnarray}
where $\Lambda_1=\beta_1+c$, $\Lambda_2=\pi+\beta_1+c$,
$\Lambda_3=\beta_1-c$, $\Lambda_4=\pi+\beta_1-c$. Consider the
term with $\Lambda_1$. Derivative w.r.t. $\Lambda_1$ gives
\begin{equation}
\frac{\partial}{\partial\Lambda_1}F^{\Lambda_1}_{u_2
u_2}=-\frac{\beta A_+}{8\pi\beta_1} \int_{-\infty}^{\infty}dt
V'_{u_2}\frac{t}{t^2+\Lambda^2_1}.
\end{equation}
The integral can be done by simple contour integration. Then
integrating back w.r.t. $\Lambda_1$ gives the final result
\begin{equation}
F^{\Lambda_1}_{u_2 u_2} =\frac{\beta
A_+}{8\pi\beta_1}h_{\Lambda_1},
\end{equation}
where we defined
\begin{equation}
h_{\Lambda_i}=\frac{\pi}{2}\ln\left[(\Lambda_i+\beta_1)^2-c^2\right]+
\frac{\pi}
{2\beta_1}
\left[ c-\frac{a_1s^2_m}{2c}\right] \ln\frac{\Lambda_i+\beta_1+c}{\Lambda_i+
\beta_1-c}.
\end{equation}
In terms of these definitions, collecting similar terms from other
$\Lambda_i$ terms, we finally have
\begin{equation}
F_{u_2u_2}\approx\frac{\beta}{8\pi\beta_1}\left[A_+(h_{\Lambda_1}-h_{\Lambda_2})
+A_-(h_{\Lambda_3}-h_{\Lambda_4})\right].
\end{equation}
In the limit $\lambda_2\ll 1$, $A_{\pm}\rightarrow\beta_1$,
$h_{\Lambda_i}\rightarrow \pi\ln[\Lambda_i+\beta_1]$, and
$\beta_1\rightarrow x_d$, so that the metallic limit of
$F_{u_2u_2}$ is given by
\begin{equation}
F^{(met)}_{u_2u_2}\approx\frac{\beta}{4}\ln x_d,
\end{equation}
where we have omitted the constant terms.

\item Free energy $F_{u_1u_2}$:

This again uses the approximate $\sigma_{u_2}$, which is valid in
the crossover regime as well as the metallic limit. We proceed
like $F_{u_1 u_1}$. Using $V_{u_1}$ and $\Omega_{u_2}$, and doing
the $t$-integral first, we generate a diverging constant term that
we ignore. The rest becomes
\begin{equation}
F_{u_1
u_2}\approx-\frac{\beta}{4\beta_1}\int_0^{\infty}\frac{dq}{q}
\cosh (qa_0/2)e^{-(\pi
q/2)}[A_+e^{-q\Lambda_+}+A_-e^{-q\Lambda_-}]
\end{equation}
where $\Lambda_{\pm}=\beta_1\pm c$. Taking a derivative w.r.t.
$a_0$ we get rid of the $1/q$ term and replace the $\cosh(qa_0/2)$
factor with $\sinh(qa_0/2)$. The integral can then be done, which
can be integrated back w.r.t. $a_0$, giving
\begin{equation}
F_{u_1 u_2}\approx\frac{\beta}{8\beta_1}\left[A_+\ln
[(\pi+2\Lambda_+)^2-a_0^2] +A_-\ln
[(\pi+2\Lambda_-)^2-a_0^2]\right].
\end{equation}
As before, for $\lambda_d > 1$, $a_0$ is imaginary. In the limit
$\lambda_2\ll 1$, we get
\begin{equation}
F^{(met)}_{u_1u_2}\approx\frac{\beta}{4}\ln x_d.
\end{equation}

\item Free energy $F_{2\Gamma}$:

Since $V_2$ is already only up to order $\lambda_2$, this term is
only valid in the metallic limit.  We use the $\lambda_2
\rightarrow 0$ limit for $\sigma_{\Gamma}$.
\begin{equation}
F_{2\Gamma}=\frac{\Gamma\beta}{2}\left(V_{2\Gamma}+V_{2z}\right).
\end{equation}
The first term $F^{\Gamma}_{2 \Gamma}$ involves an integral over
$A(t)$ defined in (A20) which can be easily done, and using
$\sum_{k=1}^{\infty}\cos (k\pi)=-1/2$, gives
\begin{equation}
F^{\Gamma (met)}_{2 \Gamma}\approx\frac{\Gamma^2 x^2_2\pi^2}{8}
\end{equation}
The second term involves several parts. The first one is
\begin{eqnarray}
J_1&=&\int_0^{\infty}\frac{a^2-3t^2}{(t^2+a^2)^3}A(t) dt,
\;\;\;a=\pi/2. \nonumber \\
&=&\sum_{k=1}^{\infty}k\cos
(k\pi)\frac{\pi^2}{2}\left[b\frac{\partial^2} {\partial
b^2}+\frac{3}{2}\frac{\partial}{\partial b}\right]
\int_0^{\infty}\frac{dt}{(t^2+b)(t^2+c)},
\end{eqnarray}
evaluated at $b=\pi^2/4$ and $c=\pi^2 k^2/4$. The result for the
integral in (B69) is $(4/\pi^2)\sum_{k=1}^{\infty}
\cos(k\pi)/(k+1)^3 $, after taking the derivatives and using
values of $b$ and $c$. The sum can be obtained in terms of Zeta
function, giving
\begin{equation}
J_1=\frac{1}{\pi^2}[3\zeta(3)-4].
\end{equation}
The second part is the same as in (B8), given by
\begin{equation}
J_2=\frac{1}{\pi^2}[\zeta(2)-1)].
\end{equation}
The third part  is of the form
\begin{equation}
J_3=\int_0^{\infty}\frac{dt}{(t^2+a^2)^3}\frac{t}{\sinh (2t)}
=\frac{1}{2}\frac{\partial^2}{\partial b^2}\int_0^{\infty}
\frac{dt}{t^2+b}\frac{t}{\sinh (2t)},
\end{equation}
which can be done using the previous results. After the
derivatives it becomes
\begin{equation}
J_3=\frac{1}{\pi^4}\left[3+4\sum_{k=1}^{\infty}\frac{(-1)^k}
{(k+1)^3}+2\sum_{k=1}^{\infty}\frac{(-1)^k}{(k+1)^2}\right]
=\frac{1}{\pi^4}[3\zeta(3)+\zeta(2)-3)].
\end{equation}
The other integrals are elementary, of the form
\begin{equation}
\int_0^{\infty}\frac{dt}{(t^2+a^2)^r}=(-1)^{r-1}\frac{\sqrt{\pi}}{2}
\frac{a^{1-2r}}{(r-1)!}\frac{\pi}{\Gamma(3/2-r)}.
\end{equation}
Collecting all the terms, we finally have
\begin{equation}
\sum_i^3\int_0^{\infty}dt
J_i(t)=
\frac{1}{\pi^2}[3\zeta(3)-4].
\end{equation}
Thus the Free energy contribution is
\begin{equation}
F^{z (met)}_{2\Gamma}\approx-\frac{z\Gamma\beta
x^2_2}{2\pi^2}[4-3\zeta(3)]
\end{equation}

\end{itemize}

\section{correction to the functional integral }
Here we give some details for the deduction of the relation (65).

As we mentioned, in general the integral (64) is of the form
$
I_{\sigma}\sim1/\sqrt{det\vert u(t,s) \vert}.
$
In the
regime $\lambda_2 \ll 1$,  we can expand $u$ as $u=u_0+\lambda_2 \delta u$,
where $u_0$ is independent of $\lambda_2$. Therefore the determinant
can be also expanded as
$
det\vert u_0 \vert\times det [1+\lambda_2 u^{-1}_0\delta u].
$
The first term is independent of $\lambda_2$, and only
contributes to the normalization. The second part can be written
as
$
\exp[Tr \ln (1+\lambda_2 u^{-1}_0\delta u)]
$
Expanding the logarithm, it is clear that the correction to the
Free energy is going to be proportional to $\lambda_2$, and the
proportionality constant is small because the dominant metallic
contributions come from the $u_0$ part and $\delta u$ is a small
correction. Since all other terms in the Free energy in the metallic 
limit are of order $\Gamma\sqrt{\lambda_2}\sim 1$,
the fluctuation corrections for the functional integral
are in general negligible.

However, our saddle point density goes to zero at the lower limit
$t=0$. It is approximately constant, equal to $\Gamma$, for large
$t$. The scale over which it changes is $t^2\sim
(\lambda_2-\lambda_1)$. In general, for the saddle point value of
$\sigma$ less than the width of the peak in $u$ vs $\sigma$, the
lower limit of $\sigma(t)=0$ will matter. For example, in the
extreme case where the saddle point $\sigma=0$, the integral over
$\sigma$ would only give 1/2 the full value. We can not evaluate
this correction explicitly, but we only need the $\lambda_1$ and
$\lambda_2$ dependent contributions, which we can estimate. Let us
consider the region of integral over $t,s$ up to some constant
$\alpha$ times $\lambda^{1/2}_d=(\lambda_2-\lambda_1)^{1/2}$,
which is the scale beyond which the $t$ dependence becomes
negligible. This is the part where the correction comes from. The
integral $I_\sigma$ will have a contribution coming from the
exponent
\begin{equation}
J_{\sigma}=\int_0^{\alpha\lambda^{1/2}_d}dt
\int_0^{\alpha\lambda^{1/2}_d}ds
\delta\sigma (t)\delta\sigma (s) u(t,s).
\end{equation}
We scale out the $\lambda_d$ from the limit by changing variables
$t^{\prime}=t/\alpha\lambda^{1/2}_d$,
$s^{\prime}=s/\alpha\lambda^{1/2}_d,$ which
leads to
$
J_{\sigma}=\lambda_d J^{\prime}_{\sigma}
$
where we defined
\begin{equation}
J^{\prime}_{\sigma}=\alpha^2\int_0^1 dt^{\prime}
\int_0^1 ds^{\prime} \delta\sigma (t^{\prime})
\delta\sigma (s^{\prime}) u(t^{\prime},s^{\prime}).
\end{equation}
Since $u$ is logarithmic in $t,s$
for small $t,s$, the change of variables will produce a
$\ln \lambda_d$ term, plus terms independent of $\lambda$.
Expand
\begin{equation}
\delta\sigma (t) =\sum_{n=-\infty}^{\infty}e^{i2\pi nt}\sigma_n;
\;\;\; u(t,s)=\sum_{m,n=-\infty}^{\infty}e^{i2\pi (mt+ns)} u_{mn},
\end{equation}
such that $J^{\prime}_{\sigma}\sim \sum_{-\infty}^{\infty}
\sigma_m u_{mn} \sigma_n$,
where
\begin{equation}
u_{mn}=\int_0^1 ds^{\prime}\int_0^1 dt^{\prime}
e^{i 2\pi (m s^{\prime}+n t^{\prime})}u(s^{\prime},t^{\prime}).
\end{equation}
It is now clear that if  $u\propto\ln \sqrt{\lambda_d}$ is the
dominant part, then only $m=0,n=0$ part is non-zero because the
integral is proportional to terms like $\sin (2\pi m)/m$. In other
words, $u_{mn}=u_0 \delta_{m,0}\delta_{n,0}$, where $u_0=-\ln
\sqrt{\lambda_d}$. Then we can write
\begin{equation}
I_{\sigma}=\int_{-\sigma_{sp}\lambda^{1/2}_d}^{\infty} d\sigma_0
e^{-\lambda_d u_0 \sigma^2_0} \sim \frac{1}{\sqrt{u_0\lambda_d}},
\end{equation}
since $\sigma_{sp}\approx 0$. (The correction will again be powers of
$\lambda$).
In the Free energy, this gives a contribution (Eq. 65)
\begin{displaymath}
\delta F\approx\frac{1}{2}\ln \lambda_d.
\end{displaymath}
Note that because of the logarithm, the coefficient of the
$\ln\lambda_d$ term is one-half, despite all the approximations.
The term appears simply from the scaling of the original variable,
which was determined by the form of the diverging part of
$\sigma_{u_1,u_2}$ which sets the scale over which the saddle
point density becomes independent of $t$ near the origin. Note
also that for $\lambda_d \gg 0$, the density no longer has a dip
at $t=0$, and the correction vanishes. Thus it will be relevant
only in the metallic limit.


\end{document}